\documentclass[
aip, singlecolumn,
amsmath,amssymb,
preprint,%
]{revtex4-1}

\usepackage{graphicx}
\graphicspath{ {Figures/} }
\usepackage{bm}

\usepackage[utf8]{inputenc}
\usepackage[T1]{fontenc}
\usepackage{mathptmx}
\usepackage{etoolbox}

\usepackage{array}
\usepackage{amsthm}
\usepackage{amsfonts}
\usepackage{enumitem}
\usepackage{bbm}
\usepackage[title]{appendix}
\usepackage{mathrsfs}
\usepackage{verbatim}
\usepackage{upgreek}
\usepackage{dsfont}
\usepackage{bbold}
\usepackage{multirow}
\usepackage{physics}
\usepackage{mathtools}
\usepackage{url}
\usepackage[colorlinks]{hyperref}

\newcommand{\ti}[1]{\textit{#1}}

\newcommand{\be}{\begin{equation}}
\newcommand{\ee}{\end{equation}}
\newcommand{\bes}{\begin{equation*}}
\newcommand{\ees}{\end{equation*}}
\newcommand{\f}[2]{\frac{#1}{#2}}

\newcommand{\parallelsum}{\mathbin{\!/\mkern-5mu/\!}}
\makeatletter
\newcommand{\vast}{\bBigg@{3}}
\newcommand{\Vast}{\bBigg@{5}}
\makeatother

\makeatletter
\def\@email#1#2{%
 \endgroup
 \patchcmd{\titleblock@produce}
  {\frontmatter@RRAPformat}
  {\frontmatter@RRAPformat{\produce@RRAP{*#1\href{mailto:#2}{#2}}}\frontmatter@RRAPformat}
  {}{}
}%
\makeatother

\begin{document}


\title[]{Tunable spin-charge conversion in class-I topological Dirac semimetals}
\author{Rui-Hao Li}
\thanks{These authors contributed equally.}
\affiliation{Department of Physics, Case Western Reserve University, Cleveland, Ohio 44106, USA}
\author{Pengtao Shen}%
\thanks{These authors contributed equally.}
\affiliation{Department of Physics, Case Western Reserve University, Cleveland, Ohio 44106, USA}
\author{Steven S.-L. Zhang}
\thanks{Electronic mail: shulei.zhang@case.edu}
\affiliation{Department of Physics, Case Western Reserve University, Cleveland, Ohio 44106, USA}

\date{\today}

\begin{abstract}
We theoretically demonstrate that class-I topological Dirac semimetals (TDSMs) can provide a platform for realizing both electrically and magnetically tunable spin-charge conversion. With time-reversal symmetry, the spin component along the uniaxial rotation axis ($z$-axis) is approximately conserved, which leads to an anisotropic spin Hall effect---the resulting spin Hall current relies on the relative orientation between the external electric field and the $z$-axis. The application of a magnetic field, on the other hand, breaks time-reversal symmetry, driving the TDSM into a Weyl semimetal phase and, consequently, partially converting the spin current to a charge Hall current. Using the Kubo formulas, we numerically evaluate the spin and charge Hall conductivities based on a low-energy TDSM Hamiltonian together with the Zeeman coupling. Besides the conventional tensor element of the spin Hall conductivity $\sigma_{xy}^z$, we find that unconventional components, such as $\sigma_{xy}^x$ and $\sigma_{xy}^y$, also exist and vary as the magnetic field is rotated. Likewise, the charge Hall conductivity also exhibits appreciable tunability upon variation of the magnetic field. We show that such tunability---as well as large spin-charge conversion efficiency---arises from the interplay of symmetry and band topology of the TDSMs.
\end{abstract}

\maketitle

\section{\label{sec:intro} Introduction}
All along, spin-charge conversion has been a major theme of spintronics. On one hand,  efficient control of magnetic states can be achieved electrically by injecting a spin current into a magnetic system; on the other hand, the transition from one magnetic state to another alters the spin current passing through the system, which can, in turn, be detected indirectly by converting it to a charge current. 

One celebrated example of spin-charge conversion is the spin Hall effect~\cite{DYAKONOV71PLA_spinHall, Dyakonov71JETP_SHE, Hirsch99PRL_SHE, sZhang00PRL_SHE, Sinova04_SHE, Vignale10_SHE, aHoffmann13ieee_SHE, Sinova15RMP_SHE}---the generation of a pure spin current propagating perpendicularly to the applied charge current---and its Onsager reciprocal, the inverse spin Hall effect~\cite{Saitoh11JAP_ISHE}. For both effects, the conversion efficiency is commonly characterized by the spin Hall angle: the ratio of the induced transverse spin (charge) current to the applied charge (spin) current. Following intensive efforts on quantifying the spin Hall angle of transition metals and their alloys~\cite{aHoffmann13ieee_SHE}, studies have also been extended to explore other materials systems~\cite{Sanchez13NatComm_SC-conv,Saitoh14PRL_SC-TI, Sanchez16PRL_SCC-a-Sn,ySun16PRL_SHE-WSM,ySun17PRB_SHE-DNL,Han18NPJ-QM_SC-review, Derunova19_ISHE-SC, Vaz19NM_sc-STO3, sZhang19PRL_sc-mWSM, luqiao20NC_sc-VO2phase, Yen&Guo20PRB_SHE_SNE-DSM, Katsuhsia20PRB_DSM-SHE, Farzaneh20PRM_ISHE-TI} for more tunable spin-charge conversion, which is potentially useful when it comes to endowing spintronic devices with new functionalities.

The geometry of spin currents generated by the spin Hall effect generally hinges on crystal symmetry. In systems with cubic symmetry such as transition metals Pt and Pd, the flow direction of the spin current, the direction of the charge current, and the spin polarization are mutually orthogonal. Moreover, the spin Hall conductivity (SHC) is typically isotropic in these materials~\cite{slZhang14EPL}. Recently, it has been shown, both theoretically and experimentally, that anisotropic spin-charge conversion with unconventional SHC components occurs in layered Weyl semimetals---such as WTe$_2$ and MoTe$_2$---due to reduced crystal symmetry~\cite{MacNeill17NP_SOT-sym-WTe2, Xue20PRB_staggered_SHE, Zhao20PRR_CSC_WTe2, Song20nmat_MoTe2_SHE}. On the other hand, magnetic materials with spin-orbit coupling allow for even more spin Hall current configurations due to the breaking of additional mirror and rotational symmetries by the magnetization~\cite{Amin19PRB_SHE_FM, Davidson20PLA_FMspin}.

In this work, we will demonstrate theoretically that the class-I topological Dirac semimetals (TDSMs)~\cite{zWang12PRB_Na3Bi, zWang13prb_Cd3As2, bjYang14NC_type-TDSM}, which possess a pair of Dirac points on a uniaxial rotation axis (usually specified as the $z$-axis), can provide another promising platform for realizing both electrically and magnetically tunable spin-charge conversion due to the interplay of symmetry and band topology. Without an external magnetic field, the TDSMs can generate a pure spin Hall current driven by an electric current. The application of a magnetic field breaks time-reversal symmetry and consequently leads to a charge Hall current---thus realizing spin-to-charge conversion---together with an unconventional spin Hall current whose spin polarization is in general not perpendicular to both the current flow direction and the electric field. In the weak-field limit, we find that the dependencies of the spin and charge Hall currents on the electric and magnetic fields can be summarized as follows
\begin{subequations}\label{eq1}
\begin{align}
	&\vb Q^{x(y)} \approx \chi^{x(y)} B_{z}B_{x(y)}(\vb E\cross \vu z), \label{eq1a}\\
	&\vb Q^z \approx \qty[\sigma^0_\text{SH}+\qty(\chi^z_\perp B^2_\perp + \chi^z_{\parallel} B_z^2)](\vb E\cross\vu z),\label{eq1b}\\
	&\vb j \approx \kappa B_z(\vb E\cross \vu z), \label{eq1c}
\end{align}
\end{subequations}
where $B^2_\perp = B_x^2+B_y^2$, the superscripts $x,\ y$, and $z$ denote the direction of spin polarization, $\chi^x,\ \chi^y,\ \chi^z_\perp,\ \chi^z_\parallel$, and $\kappa$ are the field-independent response functions, and $\sigma^0_\text{SH}$ is the intrinsic SHC without a magnetic field. A few remarks are immediately in order. 

First of all, when an electric field $\vb E$ is applied without a magnetic field, a spin Hall current, which takes the form $\mathbf{Q}^z=\sigma_\text{SH}^0\mathbf{E}\cross \vu{z}$,~\cite{aBurkov16PRL_TDSM-SHE, Katsuhsia20PRB_DSM-SHE} is generated. One notable difference from the spin Hall effect in non-topological materials such as heavy metals is that in a TDSM, only the spin current with spin polarized in the $z$-direction is generated, irrespective of the electric field direction. Furthermore, the spin Hall effect is anisotropic: it is maximized when $\vb{E}$ is orthogonal to the high symmetry $z$-axis and vanishes when $\vb{E}$ is aligned with the $z$-axis. As a result, the spin current is confined in the $xy$-plane. The same anisotropy carries over even when a magnetic field is applied, indicating that this feature is associated with the unique topology of the TDSMs rather than symmetry, as we will explain in detail in Sec.~\ref{sec3:SCC}. In this case, an electrically tunable spin current can be attained in a TDSM via the intrinsic spin Hall effect.

Similar to the magnetization in ferromagnets~\cite{Amin19PRB_SHE_FM}, the application of an external magnetic field $\vb B$ breaks time-reversal symmetry, along with additional mirror symmetries about the planes containing $\vb B$ and rotational symmetries about the axes perpendicular to $\vb B$, hence allowing for spin currents with spin polarized in directions different from the $z$-axis, such as $\vb Q^x$ and $\vb Q^y$, as given by Eq.~\eqref{eq1a}. In the meantime, a charge Hall current is now allowed by symmetry, which also varies with the magnetic field, realizing spin-to-charge conversion. As highlighted in Sec.~\ref{sec:effHam}, the phase spectrum of a TDSM is highly dependent on the direction of the magnetic field. Consequently, by varying the strength and direction of the magnetic field, which effectively changes the spin character of the bands as well as the configuration of the Weyl points split out from the Dirac points, one can modulate the spin-charge conversion in a TDSM, as we will show in detail in Sec.~\ref{sec3}.

\section{\label{sec:effHam}Low-energy effective Hamiltonians}
\begin{figure}[!t]
\centering
  \includegraphics[width=\linewidth]{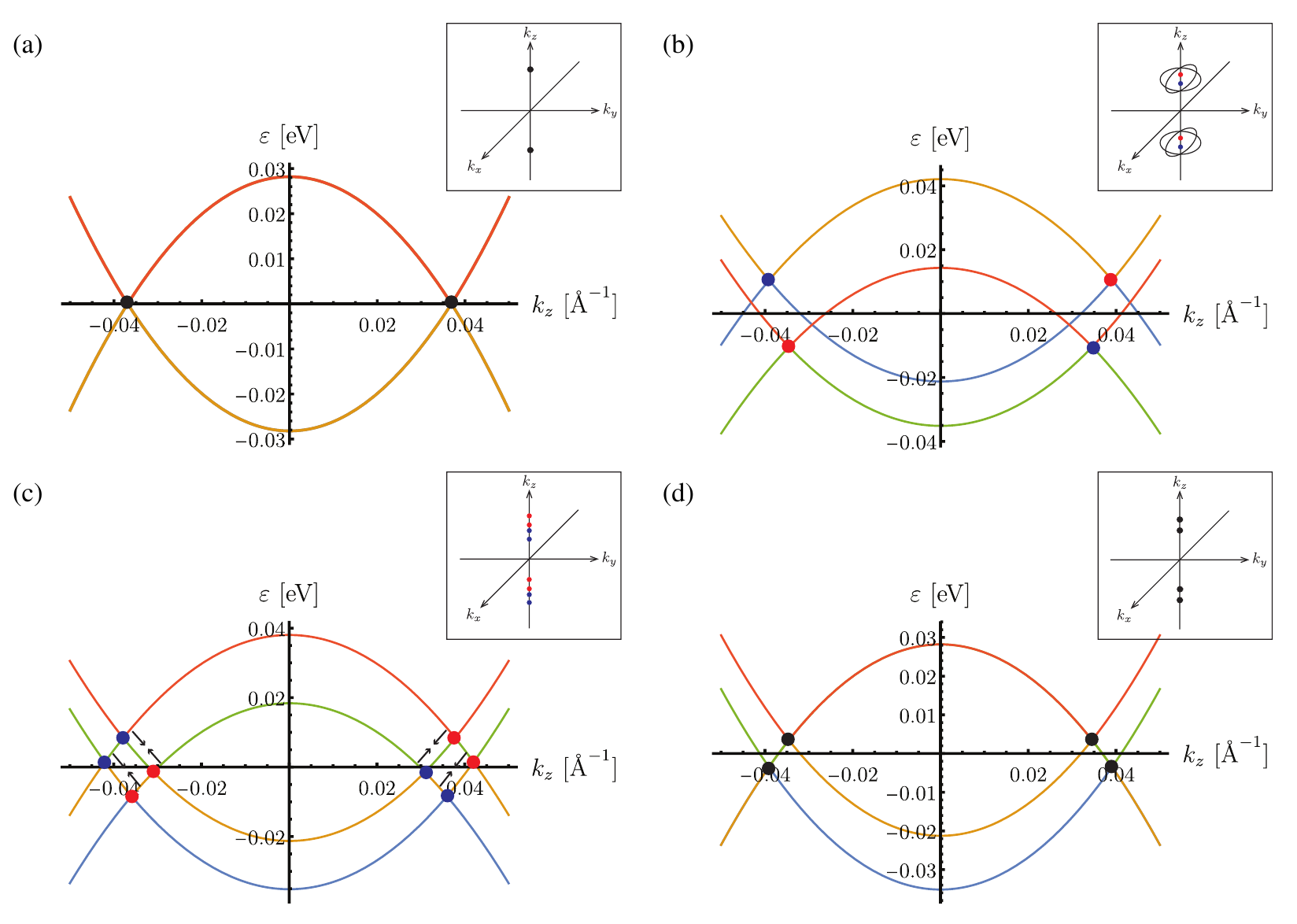}
    \caption{Evolution of the band dispersion $\epsilon_k(0,0,k_z)$ with external magnetic field : (a) $\mathbf{B}=0$, (b) $\mathbf{B}=B(0,0,1)$, (c) $\mathbf{B}=\frac{B}{\sqrt{2}}(1,0,1)$, and (d) $\mathbf{B}=B(1,0,0)$. Dirac points are indicated by black dots, while Weyl points of positive and negative chirality are denoted by blue and red dots, respectively. Insets display positions of the Dirac/Weyl points and the nodal lines in the 3D momentum space. The arrows in (c) represent the directions of movement of the Weyl points as the magnetic field is rotated from [001] to [100] direction. The parameters used here correspond to the ones for a Cd$_3$As$_2$ thin film~\cite{Arribi20PRB_CdAs_params}: $A = 1.116$~eV \AA, $M_0 = -0.0282$~eV, $M_1 = -20.72$~eV \AA$^2$, $M_2 = -13.32$~eV \AA$^2$, $\tilde g = 30$, and $B = 2$~T.} 
    \label{fig:1}
\end{figure}

Let us commence with the low-energy Hamiltonian around the $\Gamma$ point of a class-I TDSM (such as Na$_3$Bi or Cd$_3$As$_2$)~\cite{zWang12PRB_Na3Bi, zWang13prb_Cd3As2},  
\be \label{eqHD}
H_D(\vb k) =A(k_x \sigma_z\tau_x - k_y\tau_y)+m(\vb k)\tau_z+\delta H(\vb k),
\ee
where $\sigma_i$ and $\tau_i$ ($i = x,y,z$) are the Pauli matrices acting on the spin and orbital degrees of freedom respectively, the ``mass'' term can be written explicitly as 
\be
m(\vb k)=M_0-M_1 k_z^2-M_2k_{\rho}^2
\ee
with $k_{\rho}\equiv \sqrt{k_x^2+k_y^2}$, $\delta H(\vb k)=Gk_z(k_{-}^2\sigma_{+}+k_{+}^2\sigma_{-})\tau_x $ with $k_{\pm}\equiv k_x\pm i k_y$, and $A$, $M_{i}~(i=0,1,2)$, and $G$ are material-dependent parameters. When $M_0$ and $M_1$ have the same sign, the parity-even ``mass'' term gives rise to a pair of Dirac points (as shown in Fig.~\hyperref[fig:1]{1(a)}) situated at
\be
\vb k_{\pm}=(0,0,\pm k_D) \ \text{with}\ k_D\equiv\sqrt{M_0/M_1},
\ee 
both of which are four-fold degenerate as $H_D(\vb k)$ is inversion and time-reversal invariant~\footnote{For the TDSM with an effective Hamiltonian~\eqref{eqHD}, the inversion and time-reversal operators are represented by $\mathcal P = \tau_z$ and $\mathcal T = i\sigma_y \mathcal K$ (with $\mathcal K$ being the complex conjugation operator), respectively, and $H_D(\vb k)$ satisfies $\mathcal P^{-1} H_D(\vb k)\mathcal P = H_D(-\vb k)$ and $\mathcal T^{-1} H_D(\vb k)\mathcal T = H_D(-\vb k)$.}.  Also note that the high-order term $\delta H(\vb k)\left[\propto \mathcal{O}(k^3)\right]$ only leads to a small correction as far as the states in the vicinity of the Dirac points are concerned, and hence will be neglected in our analysis for simplicity unless otherwise specified. Without the $\delta H(\vb k)$ term, $\sigma_z$ commutes with $H_D$ so that the $z$-component of the spin is a conserved quantity.

When an external magnetic field is applied to a TDSM, the Zeeman coupling of the magnetic field with both the spin and orbital angular momenta is described by $H_Z = \tilde g \mu_B(\vb L+2\vb S)\cdot \vb B$~\cite{Villanova18PRB_DSM_Zeeman, Hashimoto16PRB_sc-Zeeman-DSM}, where $\mu_B$ is the Bohr magneton, $\vb L$ and $\vb S$ represent orbital and spin angular momenta, respectively, and $\tilde g$ is the effective Land\'e g-factor. We will assume $\tilde g = 30$ throughout our analysis, which is consistent with the experimentally reported g-factors for class-I TDSMs, e.g., $\tilde g \approx 37$ for Cd$_3$As$_2$~\cite{Jeon14_CdAs_g-factor} and $\tilde g \approx 20$ for Na$_3$Bi~\cite{Xiong15_NaBi_g-factor}. With the same basis used in $H_D(\vb k)$, the Zeeman coupling can be written as
\be \label{fullzm}
H_Z=\f{\tilde g\mu_B}{2}\qty[\vb B\cdot \boldsymbol{\sigma}(\tau_0+\tau_z)+2 B_z \sigma_z(\tau_0-\tau_z)],
\ee
where $\tau_0$ is the $2\times  2$ identity matrix. Before exploring the tunability of the spin-charge conversion in a TDSM via an external magnetic field, it is instructive to depict how the band structure and the corresponding topological features, as described by the total Hamiltonian $H_\text{tot}=H_D+H_Z$, evolve with the orientation of the magnetic field. 

When the magnetic field is applied in the $z$-direction ($\vb B \parallelsum \vu z$), the TDSM transforms into a Weyl semimetal due to time-reversal symmetry breaking, with the two Dirac points split into two pairs of Weyl points of opposite chirality, as shown in Fig.~\hyperref[fig:1]{1(b)}. The $z$-component of the spin, in this case, remains conserved, so the spin polarizations of the four bands are well defined with respect to the magnetic field: The crossing points of the two spin-up bands (indicated by the yellow and blue curves) constitute a pair of Weyl points, and the two spin-down bands (indicated  by the red and green curves) also intersect each other, giving rise to the other pair of Weyl points. The energy of the two pairs is shifted in opposite directions due to the Zeeman splitting. One can also see this by separating the full Hamiltonian $H_\text{tot}$---with $\sigma_z$ replaced by its eigenvalues $s_z=\pm 1$---into two $2\times2$ block matrices (or the spin-up and -down sectors), each describing a magnetic Weyl semimetal with a pair of Weyl points. In addition, there exist four band-crossing nodal lines on the $k_x$-$k_z$ and $k_y$-$k_z$ planes, as shown in the inset of Fig.~\hyperref[fig:1]{1(b)}. Note that if the cubic term $\delta H(\vb k) \neq 0$, the nodal lines would be gapped out except on the $k_z$ axis, leaving two additional pairs of double Weyl points with Chern numbers $\pm 2$~\cite{Villanova18PRB_DSM_Zeeman, Baidya20PRB_Zeeman_CdAs}.

When the magnetic field is tilted away from the $z$-axis, the spin-dependent band structure is altered accordingly. The $z$-component of the spin is no longer conserved, as $[\sigma_z,H_D]\neq 0$, and hence the bands become spin-mixed in general. Meanwhile, the evolution of the band structure, following the change of the magnetic field orientation, is also reflected in the change of the Weyl-point configuration, which is quite notable. In particular, the nodal lines are gapped out, leading to another two pairs of Weyl points with Chern numbers $\pm 1$, as illustrated in Fig.~\hyperref[fig:1]{1(c)} with $\vb B$ pointing in [101] direction. As the magnetic field is tilted towards the $xy$-plane, the Weyl points are moving towards each other (indicated by the black arrows in Fig.~\hyperref[fig:1]{1(c)}). Finally, when the magnetic field is parallel to $xy$-plane ($\vb B \perp \vu z$), as shown in Fig.~\hyperref[fig:1]{1(d)}, each of the Weyl points that exist when $\vb B \parallelsum \vu z$ merges with a new Weyl point that is generated from gapping out of the nodal lines. Since they carry opposite Chern numbers, they form a Dirac point again. In this case, the system enters a topologically trivial Dirac semimetal phase as a result of the breaking of the mirror symmetry $\mathcal M_z$ about the $xy$-plane, in which case a topological invariant (the mirror Chern number) cannot be defined on this plane~\cite{bjYang14NC_type-TDSM}. Indeed, one can verify that with $\mathcal M_z = i\sigma_z \tau_0$, $\mathcal M_z^{-1}H_Z(B_x, B_y, 0)\mathcal M_z\neq H_Z(B_x,B_y, 0)$.

\section{\label{sec3:SCC}Spin-charge interconversion}
\label{sec3}

\subsection{\label{sec3a:zeroB}No magnetic field}
As mentioned earlier, the unique spin and topological characters of the TDSM band structure lead to distinctive features in the interconversion between charge and spin currents, which are summarized in Eq.~\eqref{eq1}. In this section we will elaborate on this. Let us first consider the TDSM Hamiltonian [Eq.~\eqref{eqHD}] with $\delta H(\vb k) = 0$ as two copies of the Weyl Hamiltonian associated with the two eigenvalues of $\sigma_z$ (i.e., $s=\pm 1$ corresponding to the spin-up and -down sectors, respectively): 
\be
H_{W}^s(\vb k) =A(s k_x \tau_x - k_y\tau_y)+m(\vb k)\tau_z,
\ee
each describing a pair of Weyl points that serve, respectively, as the source and drain of a spin-dependent Berry curvature in the $\vb k$-space, which can be expressed as 
\be \label{eqbc}
\boldsymbol{\Omega}_{n}^s(\vb k)=-n\frac{s M_1 A^2}{2\abs{\epsilon_n(\vb k)}^3}\mqty(2k_xk_z, & 2k_yk_z, & k_z^2-k_D^2-\frac{M_2}{M_1}k_{\rho}^2),
\ee
where $n = \pm$ is the index of the energy band with $\epsilon_{n}(\vb k)=n\sqrt{A^2 k_{\rho}^2+[m(\vb k)]^2}$, and $k_D = \sqrt{M_0/M_1}$ is the separation of the Weyl points on the $k_z$-axis.  

When an electric field $\vb E$ is applied,  a Weyl fermion acquires an anomalous velocity~\cite{Karplus54, Xiao2010} of the form $\tilde{\vb v}_{n}^s(\vb k) = \f{e}{\hbar}\vb E \times \boldsymbol{\Omega}^s_{n}(\vb k)$, which satisfies $\tilde{\vb v}_{n}^s(\vb k)=-\tilde{\vb v}_{n}^{-s}(\vb k)$. As a result, Weyl fermions with opposite spins are deflected in opposite directions perpendicular to the applied electric field, leading to an intrinsic spin Hall current with spin polarized in the $z$-direction that is given by $\vb Q^z= \sum_{s,n} s \int_{\vb k} \tilde{\vb v}_{n}^{s}(\vb k) f^0_{n\vb k}$, where $\int_{\vb k}$ is a shorthand notation for $\int d^3\vb k/(2\pi)^3$, and $f^0_{n\vb k}$ describes the equilibrium Fermi-Dirac distribution of the Weyl fermions [at zero temperature $f^0_{n\vb k} = \Theta(\mu-\epsilon_{n}(\vb k))$ with $\mu$ being the chemical potential and $\Theta(x)$ denoting the Heaviside step function]. On the other hand, the anomalous Hall current $\vb j = \sum_{s,n} \int_{\vb k} \tilde{\vb v}_{n}^{s}(\vb k) f^0_{n\vb k}$ vanishes. Carrying out the integration of the spin Hall current, we find~\footnote{In general topologically trivial bands near the Fermi energy may also have sizable contributions to the spin Berry curvature and modify the spin Hall responses if band degeneracy occurs. However, in the two prototypical class-I TDSM materials, Cd$_3$As$_2$ and Na$_3$Bi, the pair of Dirac points are close to the Fermi energy with very few trivial bands nearby. Therefore, the low-energy electronic properties as well as the intrinsic SHC in these two materials are well captured by the effective four-band Hamiltonian \eqref{eq1}.}
\be \label{jzeqn}
\vb Q^z =\sigma_\text{SH}^0 (\vb E \times \vb z),
\ee
where $\sigma_\text{SH}^0$ characterizes the magnitude of the intrinsic spin Hall conductivity $\sigma^z_{xy}$ in the TDSM when there is no magnetic field present. Clearly, the magnitude of the spin Hall current reaches its maximum when the electric field is perpendicular to the $z$-axis. Moreover, the SHC is proportional the separation between the Dirac points (which reside right on the Fermi level) and takes a neat form 
\be \label{SH0}
\sigma_\text{SH}^0 = \frac{e k_D}{\pi^2\hbar},
\ee
where $-e$ is the electron charge and $2k_D$ is the distance between the Dirac points on the $k_z$-axis. This is analogous to the anomalous Hall conductivity in a Weyl semimetal with a pair of Weyl points, where it is proportional to the distance between the Weyl points in the reciprocal space~\cite{Burkov14PRL_AHE-WSM}.

\begin{figure}[!t]
\includegraphics[width=0.6\linewidth]{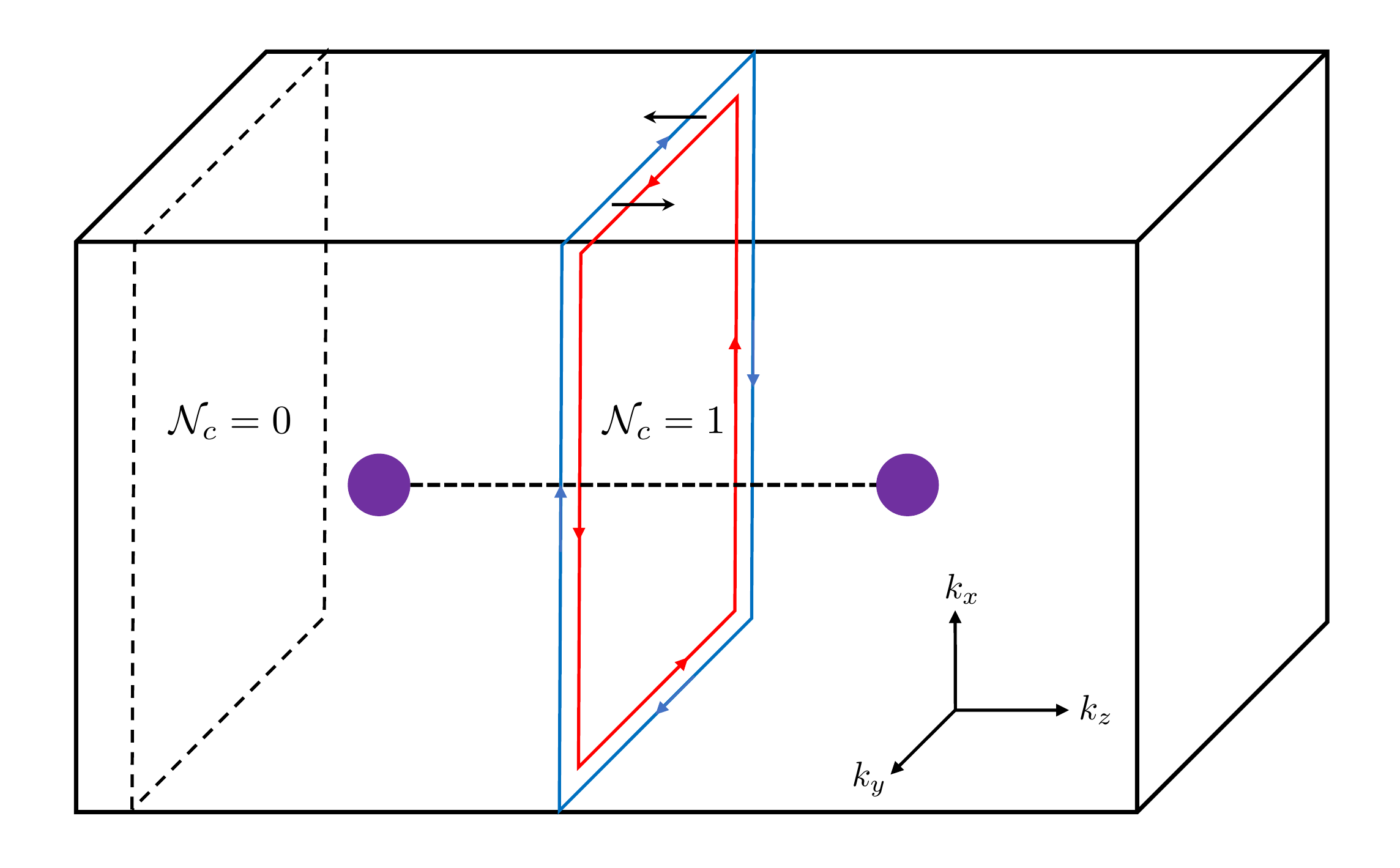}	
\caption{Two Dirac points shown in purple circles are situated at $\qty(0,0,\pm k_D)$ in reciprocal space. Each $k_x$-$k_y$ plane with a fixed $k_z\in (-k_D, k_D)$ is a 2D quantum spin Hall insulator characterized by a $\mathbb Z_2$ topological charge $\mathcal N_c = 1$. The quantum spin Hall insulator hosts 1D helical edge states carrying spin Hall currents that flow in the $k_x$-$k_y$ plane (blue and red arrows), with spin polarized in the $k_z$-direction (black arrows). Any 2D plane with a $k_z$ outside $(-k_D, k_D)$ is a topologically trivial insulator with $\mathcal N_c = 0$.}
\label{fig:SHEtopo}
\end{figure}

As pointed out in Sec.~\ref{sec:intro}, there are two remarkable and interrelated features associated with Eq.~\eqref{jzeqn}---that the spin is polarized solely in the $z$-direction and that the spin current always flows perpendicular to the $z$-axis, regardless of the electric field direction---both of which are intimately tied to the topology of a TDSM. It is not hard to see from Eq.~\eqref{eqbc} that only the $z$-component of the Berry curvature is responsible for the spin Hall current in a TDSM, as the $x$- and $y$-components are odd in $k_x$ and $k_y$, respectively. The $z$-component of the Berry curvature is linked to a $k_z$-dependent $\mathbb{Z}_2$ topological charge given by
\be
\mathcal{N}_c(k_z)\equiv \frac{1}{4\pi}\sum_s s\int dk_{x}dk_{y} \boldsymbol{\Omega}_{-,z}^s(\vb k)=\Theta\left(k_D^2-k_z^2\right),
\ee
where we have only considered the occupied band with $n=-1$ and $M_{1,2}<0$. It is clear that this $\mathbb Z_2$ charge is nonzero only for $k_z\in (-k_D,k_D)$. Essentially, the Hamiltonian $H_D$ in each $k_x$-$k_y$ plane with a fixed $k_z$ between the two Dirac points represents a 2D quantum spin Hall insulator~\cite{Hasan_Kane_10RMP_TI} with 1D helical edge states carrying the spin Hall current, whereas any 2D plane outside $(-k_D,k_D)$ carries a zero topological charge and hence is topologically trivial, as shown in Fig.~\ref{fig:SHEtopo}. The helical edge states can only carry spin currents that flow in the $xy$-plane, with spin polarization perpendicular to the plane, that is, in the $z$-direction. It follows that the SHC in a 3D TDSM can be interpreted as a collective contribution from the assembly of quantum spin Hall insulators, i.e., $\sigma_\text{SH}^0=\frac{e}{2\pi^2\hbar}\int^{+\infty}_{-\infty} dk_z \mathcal{N}_c(k_z)$. This therefore indicates that the electric tunability of the SHC manifest in Eq.~\eqref{jzeqn} is connected to the band topology rather than symmetry of a TDSM and therefore would not be altered even if additional symmetries of the system are broken by, for example, a magnetic field.

Now if one breaks time-reversal symmetry by applying an external magnetic field, the spin Hall current given by Eq.~\eqref{jzeqn} would be partially converted to a charge current, resulting in finite spin and charge Hall conductivities. As discussed in Sec.~\ref{sec:effHam}, placing a TDSM in a magnetic field changes the topological and spin characters of its band structure. Therefore, both the spin and charge Hall effects would be sensitive to the magnetic field direction, giving rise to a magnetically tunable spin-charge conversion. In the following subsections, we will discuss their magnetic-field dependence with an electric field fixed in the $y$-direction.

\subsection{Magnetic field along the $z$-axis}

 Again, we first consider a magnetic field collinear with the $z$-axis, a relatively simple case in the sense that $H_\text{tot} = H_D+H_Z$ remains block-diagonal. In this case, the Zeeman coupling [given by Eq.~\eqref{fullzm}] can be decomposed into an orbital-symmetric and an orbital-antisymmetric term, 
\be \label{zmz}
H_Z=h_{+} \sigma_z\tau_0+h_{-} \sigma_z\tau_z
\ee
with $h_+ = -3h_- = 3\tilde g \mu_B B_z/2$. While neither of the two Zeeman terms mix the spin-up and spin-down sectors, the generated Weyl points turn out to be present at different momenta and energies (as shown in Fig.~\hyperref[fig:1]{1(b)}), which makes a fully analytical solution of the spin and charge Hall conductivities intractable. However, it is still instructive to examine the effect of each Zeeman term individually, from which we can extract some information about the magnetic-field dependence of the conductivities in the perturbative regime.

Should the orbital-antisymmetric Zeeman term $H_Z^\prime =h_{-} \sigma_z\tau_z$ stand alone, the resulting energy dispersions are given by ${\epsilon_{\pm}^s}^\prime(\vb k) = \pm\sqrt{A^2 k_\rho^2+[m(\vb k)+s h_-]^2}$, where $s=\pm 1$ denotes the spin of the energy bands. Therefore, the two Dirac points would split into four Weyl points, all having the same energy $\epsilon=0$ and being situated on the $k_z$-axis at $(0,0,\pm k_{s}^\prime)$ with $k_s^\prime=\sqrt{k_D^2+ s \f{h_{-}}{M_1}}$. The resulting SHC at $\mu=0$ is given by (see App.~\ref{app:analytical_soln} for a detailed derivation) 
\be \label{shcasz}
{\sigma^{z}_{xy}}^\prime = \f{e(k'_+ + k'_-)}{2\pi^2\hbar}
\simeq \left(1-\frac{h_{-}^2}{8M^2_0}\right)\sigma_\text{SH}^0,
\ee
where we have kept terms up to $\mathcal{O}\left(B_z^2\right)$, and $\sigma_\text{SH}^0$ is given by Eq.~\eqref{SH0}. To leading order, the correction to the standard SHC that this Zeeman term brings about is proportional to $B_z^2$ and only relies on the ratio of the magnitude of the corresponding Zeeman splitting energy, $h_{-}$, to the magnitude of the Lifshitz transition point $M_0$ (at which the two low-energy Dirac cones of the TDSM merge). Moreover, $H_Z^\prime$ also leads to a finite charge Hall current $j_x^\prime = \sigma_{xy}^\prime E_y$, where the $\sigma_{xy}^\prime$ is the intrinsic anomalous Hall conductivity (AHC) given by  
\be \label{ahcasz}
\sigma_{xy}^\prime = \f{e(k'_+ - k'_-)}{2\pi^2\hbar}\simeq -\f{h_-}{2\abs{M_0}}\sigma_\text{SH}^0.
\ee
Comparing Eqs.~\eqref{shcasz} and \eqref{ahcasz}, it is plain to see that in the weak-field limit where $\abs{h_-/M_0}\ll 1$, the leading-order contribution to the SHC from the Zeeman splitting is $\mathcal O(B_z^2)$, while that to the AHC is $\mathcal O(B_z)$.

 On the other hand, the orbital-symmetric Zeeman term $H_Z^{\prime\prime} = h_{+} \sigma_z\tau_0$ leads to the  dispersions ${\epsilon_{\pm}^s}^{\prime\prime}(\vb k) = sh_+\pm \sqrt{A^2k_\rho^2+[m(\vb k)]^2}$. Hence, it splits the two Dirac points into two pairs of Weyl points with different energies, $\epsilon_s = s h_+$, both located at the same momenta $(0,0,\pm k_D)$. The corresponding spin Hall conductivity at $\mu=0$ can be expressed as 
\be \label{shcsz}
{\sigma^z_{xy}}^{\prime\prime}\simeq \left[1-\qty(\f{1}{24M_0^2}+\frac{M_2}{3M_0A^2})h_+^2\right]\sigma_\text{SH}^0,
\ee
where we have also expanded terms up to $\mathcal{O}(B_z^2)$ to retain a simple analytical form. Note that ${\sigma^z_{xy}}^{\prime\prime}$ clearly involves more material parameters (including $M_2$ and $A$) than its counterpart ${\sigma^z_{xy}}^{\prime}$, as the Weyl points are moved away from the Fermi surface. However, similar to ${\sigma^{z}_{xy}}^\prime$, the leading-order correction to the SHC is proportional to $B_z^2$ in the weak-field limit where $\abs{h_+/A}\ll \sqrt{\abs{M_0/M_2}}$ and $\abs{h_+/M_0}\ll 1$. Unlike $H_Z'$, $H_Z''$ alone does not lead to a finite AHC, due to the fact that the contribution of the hole pocket of the spin-up valence band is compensated by that of the electron pocket of the spin-down valence band, leading to an exact cancellation between the contributions from the two valence bands. It should be noted, however, in the presence of $H_Z'$, which shifts the Weyl point positions on the $k_z$-axis, creating asymmetric electron and hole pockets, $H_Z''$ would have a finite contribution to the AHC.

In summary, we find that when the magnetic field is applied along the $z$-axis, the orbital-symmetric part of the Zeeman term~\eqref{zmz} splits the degenerate spin-up and -down bands in a TDSM by an equal amount, generating two pairs of Weyl points at energies $\pm h_+$. Moreover, the orbital-antisymmetric part further shifts the positions of the Weyl points on the $k_z$ axis, leading to what is depicted in Fig.~\hyperref[fig:1]{1(b)}. The combination of these two terms leads to a correction to the SHC $\sigma^z_{xy}$ of $\mathcal O(B_z^2)$ and an AHC $\sigma_{xy}$ of $\mathcal O(B_z)$ to leading order, thereby realizing the spin-to-charge conversion. A more general magnetic-field dependence of the SHC and AHC can be inferred from symmetry principles, as we will show in Sec.~\ref{subsec3B} below.

\subsection{\label{subsec3B}Magnetic field in arbitrary directions}

When the magnetic field is not collinear with the $z$-axis, the spin polarization of the bands is no longer well-defined to be along the $z$-axis. Consequently, an unconventional SHC tensor---with nonvanishing $\sigma_{xy}^x$ and $\sigma_{xy}^y$ elements in particular---emerges, where the electric field, the spin and flow directions of the spin Hall current are not mutually perpendicular anymore, in contrast to the spin Hall effect in widely used heavy metals such as Pt.   

Before we proceed to compute the spin-charge conversion with a magnetic field in an arbitrary direction, it is beneficial to inspect the constraints imposed by symmetry on the SHC (AHC) tensor, which relates the spin (charge) Hall current to an applied electric field via $Q_a^i=\sigma^i_{ab}(\vb B)E_b$ [$j_a=\sigma_{ab}(\vb B)E_b$], where $a\neq b$. Note that in the presence of the magnetic field, $\sigma^i_{ab}$ is, in general, \textit{not} an antisymmetric rank-3 tensor. Since both the spin Hall current and the electric field are time-reversal symmetric, the SHC must be even in $\vb B$, i.e., $\sigma^i_{ab}(-\vb B)=\sigma^i_{ab}(\vb B)$. On the other hand, the charge current is odd under the time-reversal operation, it follows that the AHC is odd in $\vb B$, $\sigma_{ab}(-\vb B) = -\sigma_{ab}(\vb B)$. Furthermore, in the weak-field limit where the Zeeman splitting can be treated as an external perturbation, the SHC and AHC, up to $\mathcal O(\abs{\vb B}^2)$, can be expanded as 
\begin{subequations}\label{bdep}
\begin{align}
	\sigma^i_{ab}(\vb B) &=\sigma^{i(0)}_{ab}+\chi^{i}_{abcd} B_c B_d,\\
	\sigma_{ab}(\vb B) &= \kappa_{abc}B_c,
\end{align}
\end{subequations}
where $\sigma^{i(0)}_{ab}$ denotes the SHC when $\vb B = 0$, and the response functions $\chi^{i}_{abcd}$ and $\kappa_{abc}$ are independent of $\vb E$ and $\vb B$ fields and hence fully respect the symmetries of the TDSM. From the previous analysis, we know that if $\vb E$ points in the $y$-direction, the only nonzero $\sigma^{i(0)}_{ab}$ would be $\sigma^{z(0)}_{xy}$, which is given by Eq.~\eqref{SH0}. In addition to having time-reversal and inversion symmetries, the effective Hamiltonian of the TDSM [Eq.~\eqref{eqHD}] is also invariant under the following mirror reflection operations:
\begin{subequations}
\begin{align}
    \mathcal{M}_x^{-1}H_D(k_x,k_y,k_z)\mathcal{M}_x &=H_D(-k_x,k_y,k_z)\\
    \mathcal{M}_y^{-1}H_D(k_x,k_y,k_z)\mathcal{M}_y &=H_D(k_x,-k_y,k_z)\\
    \mathcal{M}_z^{-1}H_D(k_x,k_y,k_z)\mathcal{M}_z &=H_D(k_x,k_y,-k_z),
\end{align}
\end{subequations}
where the mirror reflection operators are $\mathcal{M}_x=i\sigma_x\tau_0$, $\mathcal{M}_y=-i\sigma_y\tau_z$, and $\mathcal{M}_z=i\sigma_z\tau_0$, respectively. Let us first consider the magnetic-field-induced spin current density $\Delta Q_x^i = \chi_{xyab}^i E_y B_a B_b$ with spin polarization in the $i$th-direction. With the three mirror symmetries given above, one can establish that 
\begin{subequations}
\begin{align}
	&-\Delta Q^x_x(E_y, B_x,-B_y, -B_z) = -\Delta Q^x_x(-E_y, -B_x,B_y, -B_z) = -\Delta Q^x_x(E_y, -B_x,-B_y, B_z)\,,\\
	& \Delta Q^y_x(E_y, B_x, -B_y, -B_z) = \Delta Q^y_x(-E_y, -B_x,B_y, -B_z) = -\Delta Q^y_x(E_y, -B_x,-B_y, B_z)\,, \\
	& \Delta Q^z_x(E_y, B_x, -B_y, -B_z) = -\Delta Q^z_x(-E_y, -B_x,B_y, -B_z) = \Delta Q^z_x(E_y, -B_x,-B_y, B_z)\,.
\end{align}
\end{subequations}
It then follows that the nonvanishing tensor elements of the spin Hall response functions allowed by symmetry are: $\chi_{xyxz}^x$, $\chi_{xyyz}^y$, $\chi_{xyxx}^z$, $\chi_{xyyy}^z$, and $\chi_{xyzz}^z$. This implies that when the Zeeman splitting is small, the SHC should have the following $\vb B$-dependence: 
\begin{subequations} \label{shcbdep}
\begin{align}
	& \sigma^x_{xy}(\vb B)\propto B_x B_z,\\
	& \sigma^y_{xy}(\vb B)\propto B_y B_z,\\
	& \Delta \sigma^z_{xy}(\vb B) = \alpha B_x^2 + \beta B_y^2+ \gamma B_z^2,
\end{align}
\end{subequations} 
where $\Delta\sigma^z_{xy}(\vb B) \equiv \sigma^z_{xy}(\vb B) - \sigma^{z(0)}_{xy}$ and $\qty{\alpha, \beta,\gamma}$ are proportionality constants. Similarly, imposing the mirror symmetries on the charge current $j_x = \kappa_{xya}E_y B_a$ leads to
\be
-j_x(E_y, B_x,-B_y, -B_z) = j_x(-E_y, -B_x,B_y, -B_z) = j_x(E_y, -B_x,-B_y, B_z),
\ee
from which one can conclude that the only nonvanishing tensor element is $\kappa_{xyz}$. Therefore, up to leading order, the magnetic-field dependence of the AHC is
\be \label{ahcbdep}
\sigma_{xy}(\vb B)\propto B_z
\ee
in the weak-field limit. A more generalized version of the symmetry analysis based solely on the underlying point group symmetries of crystals is presented in App.~\ref{app:symm}.

\begin{figure}[!ht]
\centering
  \includegraphics[width=\linewidth]{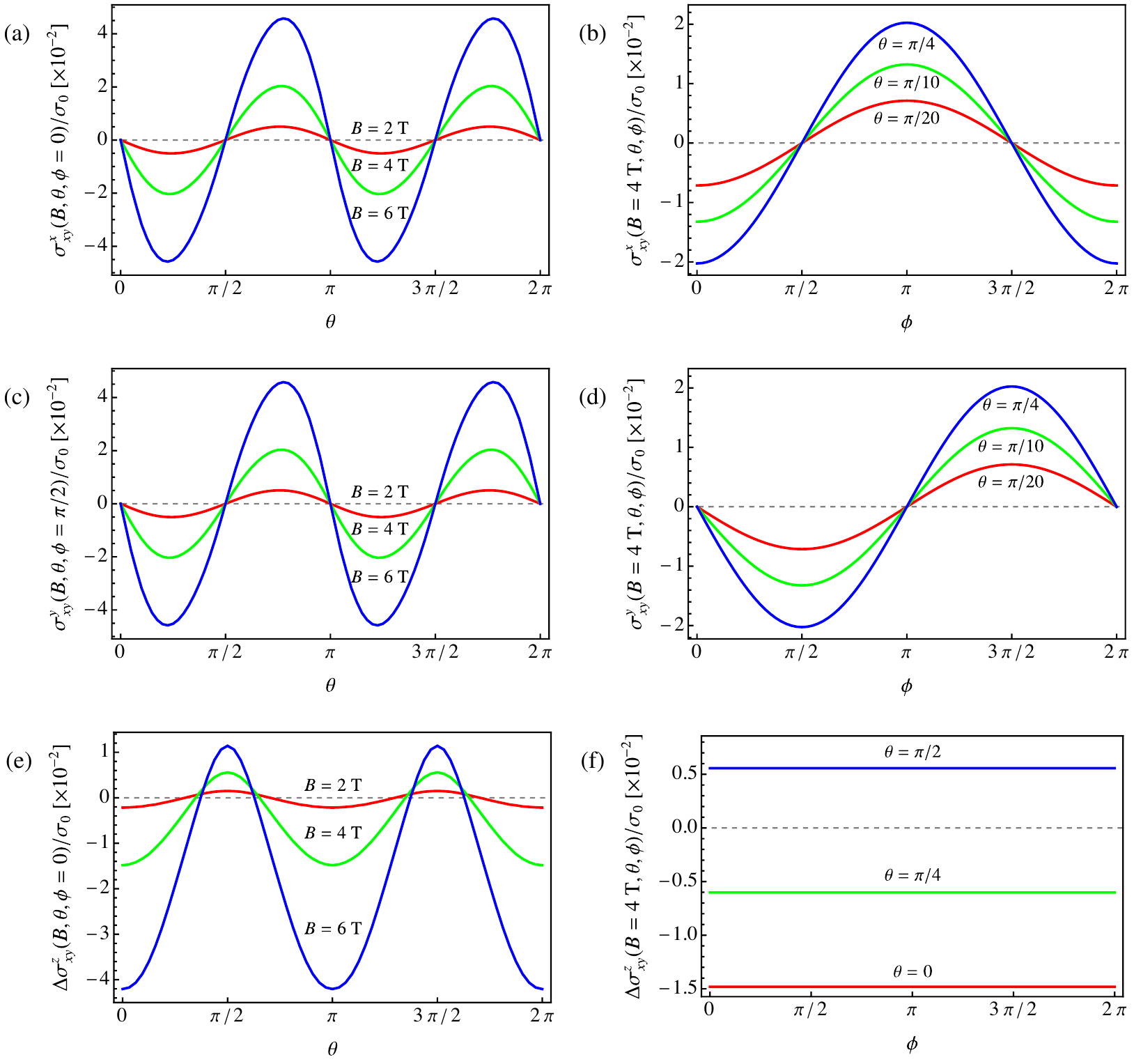}
       \caption{Dependence of the SHC's $\sigma^x_{xy}$, $\sigma^y_{xy}$, and $\Delta\sigma^z_{xy} \equiv \sigma^z_{xy}-\sigma_0$ normalized by $\sigma_0$ on the magnetic field direction, where $\sigma_0$ is the SHC in the absence of a magnetic field. $\theta$ is the polar angle of the magnetic field (relative to the $z$-axis) and $\phi$ is the azimuthal angle. (a), (c), (e) $\theta$-dependence of the SHC's at fixed $\phi$, with different magnetic field strengths $B = 2, 4, 6$ T; (b), (d), (f) $\phi$-dependence of the SHC's at a fixed magnetic field strength $B = 4$ T, with different polar angles $\theta$.} 
    \label{fig:shc}
\end{figure}

With the symmetry constraints in mind, we now proceed to calculate the SHC and AHC with a magnetic field rotated in the $xz$- and $xy$-planes to check their angular dependence. For the full Hamiltonian $H_\text{tot}=H_D+H_Z$, analytical expressions for the conductivities are no longer accessible. We therefore compute them numerically with the aid of the Kubo-Greenwood formalism for linear responses~\cite{Sinova04_SHE, Yao04_Kubo_AHC, Dutta12_Kubo_gamma, Xue20PRB_staggered_SHE}:
\begin{subequations} \label{kubo}
\begin{align}
	    \sigma_{a b}^i &= -e\hbar\int \f{d^3\vb k}{(2\pi)^3}\sum_{n}f^0_{n\vb k}\sum_{n'\neq n}\f{2\Im[\mel{n\vb k}{J^i_{a}}{n'\vb k} \mel{n'\vb k}{v_{b}}{n\vb k}]}{(\varepsilon_{n'\vb k}-\varepsilon_{n\vb k})^2+\Gamma^2}\,, \label{shckubo} \\
	    \sigma_{a b}&= -e\hbar\int \f{d^3\vb k}{(2\pi)^3}\sum_{n}f^0_{n\vb k}\sum_{n'\neq n}\f{2\Im[\mel{n\vb k}{v_{a}}{n'\vb k} \mel{n'\vb k}{v_{b}}{n\vb k}]}{(\varepsilon_{n'\vb k}-\varepsilon_{n\vb k})^2+\Gamma^2}\,, \label{ahckubo}
\end{align}
\end{subequations}
where $i,a,b=x,y$, or $z$ ($a\neq b$ for the spin and anomalous Hall effects), $\ket{n\vb k}$ is the $n$th eigenstate of the total Hamiltonian $H_\text{tot}$ with energy $\varepsilon_{n\vb k}$, and $f^0_{n\vb k} = \Theta(\mu-\varepsilon_{n\vb k})$ is the Fermi-Dirac distribution function at zero temperature. $J^i_a \equiv \f{1}{2}\qty{\sigma_i, v_a}$ is the spin-current operator with spin polarization in the $i$th direction and flow in the $a$th direction, where $\sigma_i$ is the $i$th Pauli matrix and $v_a = \f{\partial H_\text{tot}(\vb k)}{\hbar \partial k_a}$ is the velocity operator. Moreover, to approximate the effect of disorder, we have retained a constant band broadening factor that is usually omitted in the clean limit ($\Gamma \to 0$). The broadening factor is inversely proportional to the lifetime of the quasiparticles, i.e. $\Gamma \simeq \hbar/\tau$, which we estimate to be $\sim 10$~meV using a quantum scattering time $\tau \simeq 6.8\times 10^{-14}$~s measured in Cd$_3$As$_2$ thin films~\cite{Uchida17_CdAs_tauq}.

\begin{figure}[!t]
\centering
      \includegraphics[width=\linewidth]{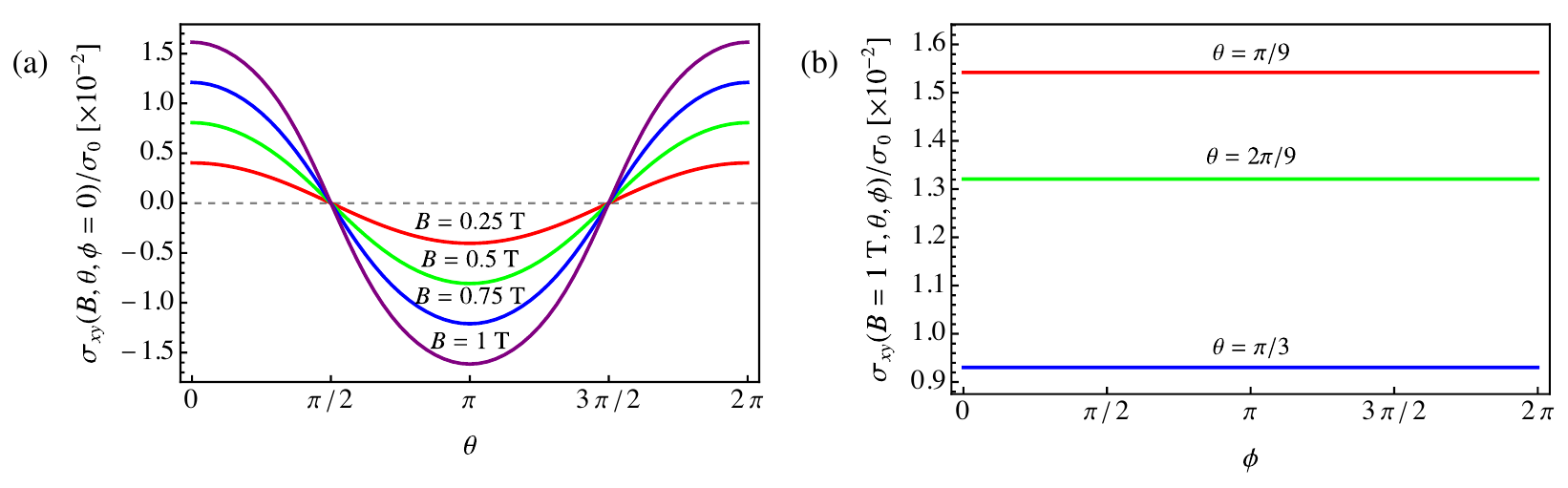}
    \caption{Dependence of the normalized anomalous Hall conductivity $\sigma_{xy}/\sigma_0$ on the magnetic field direction, where $\sigma_0$ is the spin Hall conductivity in the absence of a magnetic field. $\theta$ is the polar angle of the magnetic field (relative to the $z$-axis) and $\phi$ is the azimuthal angle. (a) $\theta$-dependence of $\sigma_{xy}$ at $\phi = 0$, with different magnetic field strengths $B = 0.25, 0.5, 0.75, 1$ T; (b) $\phi$-dependence of $\sigma_{xy}$ at a fixed magnetic field strength $B = 1$ T, with different polar angles $\theta = \pi/9, 2\pi/9, \pi/3$.} 
    \label{fig:ahc}
\end{figure}

In Figs.~\ref{fig:shc} and ~\ref{fig:ahc}, we show the numerical results of the SHC and AHC as functions of $\theta$ and $\phi$, which correspond to the polar and azimuthal angles of the magnetic field, respectively. One can see that the unconventional SHC $\sigma^x_{xy}$ and $\sigma^y_{xy}$ are generally non-vanishing when the magnetic field is applied. Moreover, both the magnitude and direction of the spin and charge Hall conductivities can be tuned by varying the direction of the magnetic field. The SHC plots in Fig.~\ref{fig:shc} show that $\sigma^x_{xy}$ and  $\sigma^y_{xy}$ have the angular dependence $\sigma^x_{xy}\sim\sin2\theta\cos\phi\sim B_x B_z$ and $\sigma^y_{xy}\sim\sin2\theta\sin\phi \sim B_y B_z$, which agree with the previous symmetry analysis given in Eq.~\eqref{shcbdep}. Furthermore, Figs.~\hyperref[fig:shc]{3(e)} and \hyperref[fig:shc]{3(f)} imply that $\Delta\sigma^z_{xy}\sim c_1\sin^2\theta+c_2\cos^2\theta$, which is independent of angle $\phi$, suggesting $\alpha=\beta$ in Eq.~\eqref{shcbdep}. However, it should be noted that this feature is not general and there would be a weak $\phi$-dependence if one includes higher order terms $\delta H$ in the Hamiltonian. Overall, combining with the $\vb E$-dependence of the spin currents discussed earlier in Sec.~\ref{sec3a:zeroB}, we arrive at the main results of the spin Hall currents given in Eqs.~\eqref{eq1a} and \eqref{eq1b}. On the other hand, the AHC plots in Fig.~\ref{fig:ahc} show that $\sigma_{xy}$ has the angular dependence $\sigma_{xy}\sim\cos\theta\sim B_z$. It is also noticeable that AHC vanishes at $\theta=\pi/2$ or $3\pi/2$ when the magnetic field is parallel to the $xy$-plane ($\vb B \perp \vu z$). This is the case when the Weyl points merge and the system enters the Dirac semimetal phase again as shown in Fig.~\hyperref[fig:1]{1(d)} . Overall, the charge Hall current has the form given by Eq.~\eqref{eq1c}. 

\begin{figure}[!t]
\centering
  \includegraphics[width=\linewidth]{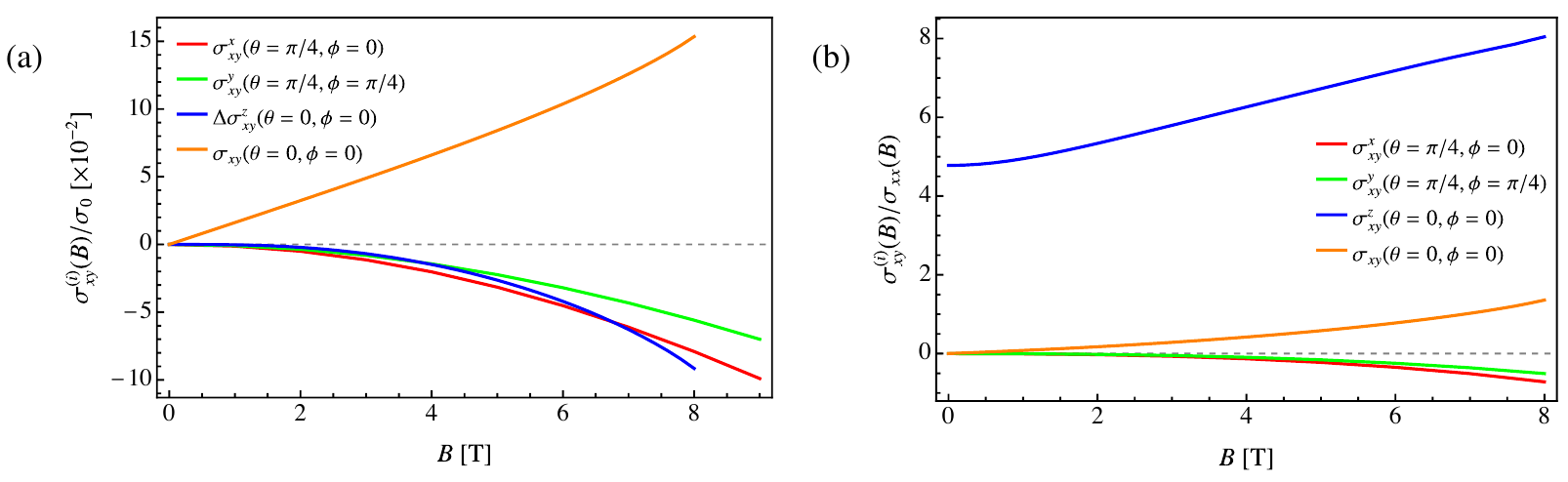}
    \caption{(a) Dependence of the normalized spin and charge Hall conductivities on the magnetic field strength with fixed magnetic field directions. (b) Dependence of spin- and charge-Hall angles on the magnitude of the magnetic field with fixed directions.} 
    \label{fig:5}
\end{figure}

Lastly, we give an estimate of the spin-charge conversion efficiency in Cd$_3$As$_2$ thin films~\cite{Arribi20PRB_CdAs_params} by plotting the dependence of the AHC and SHC on the magnetic field strength $B$ with fixed directions. The calculated spin-Hall conductivity at $B=0$ T is 9100.84/$e~\Omega^{-1}m^{-1}$, close to the measured value~\cite{Yanez_21_SCC_Cd3As2}. As displayed in Fig.~\hyperref[fig:5]{5(a)}, the calculated spin-to-charge conversion efficiency, $\sigma_{xy}/\sigma_0$, where $\sigma_0$ is the SHC in the absence of a magnetic field, is estimated to reach about $15\%$ at $B = 8$~T when the magnetic field points along the $z$-axis. In addition, one can see that the AHC and SHC show, respectively, linear and quadratic dependencies on $B$, as long as the Zeeman energy $\tilde g \mu_B B$ is lower than the Lifshitz energy $\abs{M_0}$, which is consistent with Eq.~\eqref{bdep}.

In Fig.~\hyperref[fig:5]{5(b)}, we show the theoretical predictions of the spin- and charge-Hall angles, which is defined as the ratio of the spin/charge Hall conductivities to the longitudinal conductivity, on the magnitude of the external magnetic field. The longitudinal conductivity can be separated into an interband contribution, 
\be
\sigma^\text{inter}_{xx} = e\hbar \int\f{d^3\vb k}{(2\pi)^3}\sum_{n}\sum_{n'\neq n}\qty(\f{f^0_{n\vb k}-f^0_{n'\vb k}}{\varepsilon_{n'\vb k}-\varepsilon_{n\vb k}})\f{\Gamma}{\qty(\varepsilon_{n\vb k}-\varepsilon_{n'\vb k})^2+\Gamma^2}\abs{\mel{n\vb k}{v_x}{n'\vb k}}^2,
\ee
and an intraband contribution,
\be
\sigma^\text{intra}_{xx} = e\hbar\int\f{d^3\vb k}{(2\pi)^3}\sum_n\qty(-\pdv{f^0_{n\vb k}}{\varepsilon_{n\vb k}})\f{\abs{\mel{n\vb k}{v_x}{n\vb k}}^2}{\Gamma},
\ee
where $\Gamma$ is the broadening factor introduced previously. Note that the intraband contribution corresponds exactly to the Drude conductivity in metals. From the calculations, we find that the intraband contribution is negligible at zero magnetic field and increases with increasing field strengths, due to an enhanced density of states at the Fermi energy. In contrast, the interband contribution is nonzero at zero field and decreases with an increasing magnetic field. Since the intraband contribution is two orders of magnitude smaller than its interband counterpart even at $B = 8$ T, the total longitudinal conductivity is dominated by the interband effect. This leads to increasing magnitudes of the spin- and charge-Hall angles as the magnetic field strength rises. This finding qualitatively agrees with the recent experimental results of the longitudinal magnetoresistance of Cd$_3$As$_2$ when the electric and magnetic fields are perpendicularly aligned~\cite{liang2015,li2016,li2018}. Our calculations predict that in an ideal Dirac semimetal phase with $\mu = 0$, the conventional spin Hall angle $\sigma_{xy}^z/\sigma_{xx}$ is about 4.8 at $B = 0$ and can elevate to 8 at $B = 8$ T.

\section{Summary and Outlook}
In this work, we theoretically demonstrated that the spin-charge conversion in class-I TDSMs is exquisitely sensitive to the directions of the electromagnetic fields with respect to the high-symmetry axis (or the $z$-axis), and hence possesses appreciable electric and magnetic tunability, as summarized by Eq.~\eqref{eq1}. While the electric tunability is tied to the unique band topology of TDSMs, the magnetic tunability arises from specific symmetry breaking as the magnetic field direction is varied---a TDSM may be transformed into either a WSM or a topologically trivial DSM, depending on the direction of the magnetic field relative to the $z$-axis. The resulting spin-charge conversion is thus highly anisotropic: it is minimal when the electric field is along the $z$-axis and is most efficient when the electric field is perpendicular to the $z$-axis. In the latter case, the magnetic field also induces extra spin Hall currents whose spin polarization is not perpendicular to both the current flow direction and the electric field, at variance with conventional spin Hall effect in heavy metals such as Pt or $\beta$-Ta. Our numerical calculations of the magnetic-field-dependence of the spin and charge Hall conductivities showed good agreement with what the symmetry analysis predicts in the weak-field limit. As a concrete example, the estimated spin-charge conversion efficiency for a Cd$_3$As$_2$ thin film reaches 12$\%$ for an experimentally accessible magnetic field of 8 T. 

To achieve large and anisotropic spin-charge conversion, it is desirable to search for TDSM materials with Dirac points located in the vicinity of the Fermi level, with minimal numbers of topologically trivial bands nearby. The Fermi level may be tuned by chemical doping and electrostatic gating~\cite{Nishihaya2018Sci}. In addition, as indicated by Eqs.~(\ref{shcasz})-(\ref{shcsz}), a TDSM material with a larger effective g-factor and/or smaller Lifshitz energy---the latter can be tuned under external pressure~\cite{zhang2017pressure}---is expected to exhibit a more sizable effect. 

We also would like to point out that the other class of Dirac semimetals (sometimes referred to as class-II Dirac semimetals\footnote{Some literature studies have called them type-II Dirac semimetals. However, the name ``type-II'' in the context of Weyl semimetals usually refers to Weyl semimetals with over-tilted Weyl cones. Therefore, to avoid unnecessary confusion, we refrain from such nomenclature and refer to the Dirac semimetals with single Dirac points located at the TRIM as class-II Dirac semimetals, in contrast to class-I considered in our study where the Dirac points are located on the rotation axis but not pinned to the TRIM.}) with single Dirac points at the time-reversal invariant momenta (TRIM) is also expected to exhibit the spin-charge conversion process discussed in this work due to the splitting of Dirac points into Weyl points once a magnetic field is applied. However, the dependence of the spin/charge Hall conductivities on the electromagnetic fields is likely to be different since the symmetries of this type of Dirac semimetals are generally different from those of class-I TDSMs. Moreover, it was shown theoretically that a step-function singularity arises, due to the mirror anomaly, in the AHC of class-II TDSMs when the magnetic field is rotated \cite{Burkov18_mirror_anomaly_DSM}, which is not seen in class-I TDSMs \cite{Nandy19PRB_MA_typeI_DSM}.

We plan to extend our analysis in several different directions in future works. First, energy displacement and tilting of the Dirac cones (along the $z$-direction) can, in principle, be incorporated in the current formalism by a more general four-band Hamiltonian with the additional term $\epsilon_0(\vb k) = C_0+C_1 k_z^2+C_2\qty(k_x^2+k_y^2)$~\cite{zWang12PRB_Na3Bi, zWang13prb_Cd3As2}. Using the parameters for Cd$_3$As$_2$ thin films given in Ref.~\onlinecite{Arribi20PRB_CdAs_params}, our numerical calculations suggest that, while titling and energy displacement of the Dirac cones together affect the magnitude of the spin and charge Hall conductivities, they do not alter their dependence on the electric and magnetic fields significantly, and our main result, Eq.~\eqref{eq1}, still holds in this case. However, it should be noted that in general, the spin-charge conversion in TDSMs is sensitive to the geometry of the Fermi surface, which can be varied by shifting the energy of the Dirac points and tilting the Dirac cones. As an example, it was known that the intrinsic anomalous conductivity in type-II Weyl semimetals, which have over-tilted Weyl cones, depends on the tilting parameters, in contrast to that in type-I Weyl semimetals which only depends on the separation of the Weyl points~\cite{Zyuzin2016}. We thus expect that Eq.~\eqref{SH0} should be modified similarly in the case of over-tilted Dirac cones, due to the emergence of additional electron and hole pockets at the Fermi energy. Therefore, a more systematic study of the effect of tilting and energy displacement of the Dirac cones on the spin-charge conversion in TDSMs is desired.

Second, in the present study, we considered only the Zeeman contribution of a magnetic field. Another contribution, i.e., the orbital effect, which induces the Landau levels, is expected to be important in the quantum limit with high magnetic fields and can in principle be incorporated into the effective Hamiltonian through the Peierls substitution~\cite{Liu10PRB_model_H_TI}. Notably, as demonstrated in several experiments in recent years, the chiral Landau levels in topological semimetals can conspire with the surface Fermi arcs to form closed cyclotron orbits dubbed the Weyl orbits, resulting in Shubnikov–de Haas (SdH) oscillations and 3D quantum Hall effect~\cite{Potter14ncomm_QO_TSM, Zhang17ncomm_Weyl_orbit_DSM, Zhang19nature_QHE_Cd3As2}. It is appealing to learn what roles, if any, the Weyl orbits will play in the spin-charge conversion of the TDSMs. 

Another direction to investigate the effects of the topological surface states would be to employ heterostructures composed of a magnetic material and a TDSM, wherein the magnetic proximity effect may also play an important role. On a related note, it was theoretically proposed that the surface of Weyl semimetals can host a ``hot line'' with a divergent surface Berry curvature that is distinct from the Fermi arcs~\cite{Wawrzik21PRL_hot_line_WSM}. It would be interesting to study such surface contribution as the enhanced Berry curvature could potentially lead to a large anomalous Hall response on the surface and hence a much improved spin-charge conversion efficiency in the TDSMs.

\begin{acknowledgments}
This work was supported by the College of Arts and Sciences, Case Western Reserve University. This work made use of the High Performance Computing Resource in the Core Facility for Advanced Research Computing at Case Western Reserve University. S.~S.-L.~Zhang would like to thank Olle Heinonen and Hyowon Park for helpful discussions.  
\end{acknowledgments}

\appendix
\renewcommand\thefigure{\thesection.\arabic{figure}}  

\section{\label{app:analytical_soln}Derivation of the spin and charge Hall conductivities with magnetic field along the $z$-axis}

As mentioned in the main text, when the external magnetic field is collinear with the $z$-axis, the Zeeman Hamiltonian can be written as the combination of an orbital-symmetric and an orbital-antisymmetric term. Below we provide a detailed derivation of their contributions to the spin and anomalous Hall conductivities up to the second order in $B_z$.

\subsection{Contribution of the orbital-antisymmetric Zeeman term}

With the orbital-antisymmetric Zeeman coupling $H'_Z = h_-\sigma_z\tau_z$ alone, the total Weyl Hamiltonian becomes ${H_W^s}^\prime = A(s k_x\tau_x-k_y\tau_y)+[m(\vb k)+sh_-]\tau_z$, with the corresponding dispersion given by ${\epsilon_{\pm}^s}^\prime = \pm \sqrt{A^2 k_\rho^2+[m(\vb k)+sh_-]^2}$, where $m(\vb k)\equiv M_0-M_1 k_z^2-M_2 k_\rho^2$ and $s = \pm 1$ being the spin indices of the energy bands. The Zeeman term $H'_Z$ lifts the spin degeneracy of the energy bands, generating four Weyl points located at $(0,0,\sqrt{k_D^2+s\f{h_-}{M_1}})$ with $k_D \equiv \sqrt{M_0/M_1}$, all having the same energy $\epsilon = 0$. In this case, the $z$-components of the Berry curvatures for the conduction ($+$) and valence ($-$) bands are given by
\be
{\Omega_\pm^s}^\prime(\vb k) = \mp \f{s A^2\qty(M_1 k_z^2-M_0-M_2k_\rho^2-sh_-)}{2\abs{{\epsilon_\pm^s}^\prime}^3}\,.
\ee
When an electric field $\vb E$ in the $y$-direction, a spin Hall current is induced, which can be evaluated via ${\vb Q^{z}}^\prime =\f{\hbar}{2}\sum_{s,n} s \int_{\vb k} \tilde{\vb v}_{n}^{s\:\prime} (\vb k) f^0_{n\vb k}$, with $\tilde{\vb v}_{n}^{s\:\prime} (\vb k) = \f{e}{\hbar}\vb E \times {\boldsymbol{\Omega}^s_{n}}^\prime (\vb k)$ [$n=\pm$] being the anomalous velocity. The corresponding spin Hall conductivity (SHC) can be calculated as follows
\begingroup
\allowdisplaybreaks
\begin{align*}
    {\sigma^z_{xy}}^\prime &= \f{e}{\hbar}\int\f{d^3\vb k}{(2\pi)^3}\qty[{\Omega^+_-}^\prime(\vb k)-{\Omega^-_-}^\prime(\vb k)]\\
&= \f{e}{4\pi^2\hbar}\int_{-\infty}^\infty dk_z\int_0^\infty dk_\rho k_\rho \qty[\f{A^2\qty(M_1 k_z^2-M_0-M_2k_\rho^2-h_-)}{2\qty(A^2k_\rho^2+\qty[m(\vb k)+h_-]^2)^{3/2}}+\f{A^2\qty(M_1 k_z^2-M_0-M_2k_\rho^2+h_-)}{2\qty(A^2k_\rho^2+\qty[m(\vb k)-h_-]^2)^{3/2}}]\\
&= \f{e}{8\pi^2\hbar}\int_{-\infty}^\infty dk_z \qty{\qty[1-\text{sgn}\qty(M_0-M_1k_z^2+h_-)]+\qty[1-\text{sgn}\qty(M_0-M_1k_z^2-h_-)]}\\
&= \f{e}{2\pi^2\hbar}\qty(\sqrt{\f{M_0+h_-}{M_1}}+\sqrt{\f{M_0-h_-}{M_1}})\\
&\simeq \f{e k_D}{\pi^2\hbar}\qty(1-\f{h_-^2}{8M_0^2})\,.
\end{align*}
\endgroup
This is Eq.~(12) in the main text. Note that only the completely filled valence bands contribute to the SHC, while the conduction bands are empty when the chemical potential $\mu = 0$. On the other hand, since time-reversal symmetry is broken by the magnetic field, an anomalous Hall current is also induced by the electric field, with the corresponding anomalous Hall conductivity (AHC) given by
\begin{align*}
{\sigma_{xy}}^\prime &= \f{e}{\hbar}\int\f{d^3\vb k}{(2\pi)^3}\qty[{\Omega^+_-}^\prime(\vb k)+{\Omega^-_-}^\prime(\vb k)]\\
&= \f{e}{2\pi^2\hbar}\qty(\sqrt{\f{M_0+h_-}{M_1}}-\sqrt{\f{M_0-h_-}{M_1}})\\
&\simeq -\f{ek_D}{\pi^2\hbar}\f{h_-}{2\abs{M_0}}\,,
\end{align*}
which is Eq.~(13) in the main text.

\subsection{Contribution of the orbital-symmetric Zeeman term}

If only the orbital-symmetric Zeeman term $H''_Z = h_+\sigma_z\tau_0$ is present, the Weyl Hamiltonian becomes ${H_W^s}^{\prime\prime} = A(s k_x\tau_x-k_y\tau_y)+m(\vb k)\tau_z+sh_+\tau_0$ and the corresponding dispersion is ${\epsilon_{\pm}^s}^{\prime\prime} = sh_+\pm \sqrt{A^2 k_\rho^2+m^2(\vb k)}$. This Zeeman term $H''_Z$ gives rise to two pairs of Weyl nodes with different energies $\epsilon_s = sh_+$, both located at the momenta $\qty(0,0,\pm k_D)$. The Berry curvatures are given by
\be \label{bc2}
{\Omega_\pm^s}^{\prime\prime}(\vb k) = \mp \f{s A^2\qty(M_1 k_z^2-M_0-M_2k_\rho^2)}{2\qty[A^2k_\rho^2+m^2(\vb k)]^{3/2}}\,.
\ee
Since the spin-up (-down) bands are shifted up (down) due to the Zeeman splitting, part of the spin-down conduction band is now below the Fermi level ($\mu=0$), so we need to consider its contribution to the spin and anomalous Hall effects. The SHC may be divided into two parts, i.e., ${\sigma^z_{xy}}^{\prime\prime} = {\sigma}^{\prime\prime}_\text{I}+{\sigma}^{\prime\prime}_\text{II}$ with
\begin{subequations}
\begin{align}
{\sigma}^{\prime\prime}_\text{I}&=\f{e}{\hbar}\int\f{d^3\vb k}{(2\pi)^3}\qty[{\Omega^+_-}^{\prime\prime}(\vb k)-{\Omega^-_-}^{\prime\prime}(\vb k)]\,, \\
{\sigma}^{\prime\prime}_\text{II}&=-\f{2e}{\hbar}\int\f{d^3\vb k}{(2\pi)^3}{\Omega^-_+}^{\prime\prime}(\vb k)\Theta\qty(h_+-\sqrt{A^2k_\rho^2+m^2(\vb k)})\,,
\end{align}
\end{subequations}
where $\Theta(x)$ is the Heaviside step function and we have used ${\Omega^+_-}^{\prime\prime}(\vb k) = {\Omega^-_+}^{\prime\prime}(\vb k)$. Since the orbital-symmetric Zeeman coupling does not modify the Berry curvatures of the TDSM,  $\sigma^{\prime\prime}_\text{I}$ is essentially the same as the SHC  $\sigma^0_\text{SH}$ of the TDSM in the absence of the magnetic field, which is given by Eq.~(9) in the main text. To evaluate ${\sigma}^{\prime\prime}_\text{II}$, we need to first specify the integration limits imposed by the Heaviside step function. Solving $h_+-\sqrt{A^2k_\rho^2+m^2(\vb k)}=0$ for $\xi\equiv k_\rho^2$ leads to two solutions:
\be
\xi_\pm = \f{-A^2+2M_2\eta\pm \sqrt{\qty(A^2-2M_2\eta)^2-4M_2^2\qty(\eta^2-h_+^2)}}{2M_2^2}\,,
\ee
where $\eta\equiv M_0-M_1k_z^2$. Given the material parameters for Cd$_3$As$_2$ thin films, $-A^2+2M_2\eta < 0$, so we may discard the solution $\xi_-$. To ensure that $k_\rho$ is real, we must have $\eta^2<h_+^2$ as well, which imposes the following constraints on $k_z$:
\be
-\sqrt{\f{M_0-h_+}{M_1}}<k_z<-\sqrt{\f{M_0+h_+}{M_1}},\quad \text{or}\quad \sqrt{\f{M_0+h_+}{M_1}}<k_z<\sqrt{\f{M_0-h_+}{M_1}}\,.
\ee
It follows that 
\be
	{\sigma}^{\prime\prime}_\text{II} = -\f{2e}{4\pi^2\hbar}\qty(\int_{-\sqrt{\f{M_0-h_+}{M_1}}}^{-\sqrt{\f{M_0+h_+}{M_1}}}dk_z+\int_{\sqrt{\f{M_0+h_+}{M_1}}}^{\sqrt{\f{M_0-h_+}{M_1}}}dk_z)\int_0^{\sqrt{\xi_+}} dk_\rho k_\rho\:{\Omega^-_+}^{\prime\prime}(\vb k)\,.
\ee
By evaluating the integrals and then expanding the result up to $\mathcal O(h_+^2)$, we arrive at 
\be
{\sigma}^{\prime\prime}_\text{II} = -\f{ek_D}{\pi^2\hbar}\qty(\f{1}{24M_0^2}+\f{M_2}{3A^2M_0})h_+^2\,.
\label{sigmaII}
\ee
Summing up ${\sigma}^{\prime\prime}_\text{I}$ and ${\sigma}^{\prime\prime}_\text{II}$, the total contribution of the orbital-symmetric Zeeman term to the SHC is given by 
\be
 {\sigma^z_{xy}}^{\prime\prime} \simeq \f{ek_D}{\pi^2\hbar}\qty[1-\qty(\f{1}{24M_0^2}+\f{M_2}{3A^2M_0})h_+^2]\,,
\ee
which is Eq.~(14) in the main text.

Similarly, we can also evaluate the contribution of the orbital-symmetric Zeeman term to the AHC, which turn out to be zero due to an exact calculation of the contributions from the electron- and hole-pockets as the Berry curvatures satisfy ${\Omega^+_-}^{\prime\prime}(\vb k) = -{\Omega^-_-}^{\prime\prime}(\vb k)$ and ${\Omega^+_-}^{\prime\prime}(\vb k) = {\Omega^-_+}^{\prime\prime}(\vb k)$.

\section{\label{app:symm}Symmetry analysis of the spin and charge Hall conductivities}

Since the macroscopic properties of a crystal depend only on its point group symmetries and not on those related to translations, we can constrain the response tensors associated with the AHC and SHC using a combination of the \ti{generating symmetries} of the point group~\cite{Utermohlen21PRB_tensor}, e.g., $\qty{C_4,C_{2x},P}$ for Cd$_3$As$_2$ whose point group symmetry is $D_{4h}$. Their corresponding coordinate rotation matrices are given by
\be
\vb R_{4} =\mqty(0 & -1 & 0 \\ 1 & 0 & 0 \\ 0&0&1),\qquad \vb R_{2x} = \mqty(\dmat[0]{1,-1,-1}),\qquad \vb I = \mqty(\dmat[0]{-1,-1,-1}).
\ee
The transformation rules are that $\vb v\xrightarrow{R}\vb R\vb v$ if $\vb v$ is a polar vector and $\vb v\xrightarrow{R}\det(\vb R)\vb R\vb v$ if $\vb v$ if $\vb v$ is an axial vector, where $\det(\vb R) = \pm 1$ for proper and improper rotations, respectively. On the other hand, under inversion, $\vb v\xrightarrow{I}\vb I\vb v = -\vb v$ if $\vb v$ is a polar vector and $\vb v\xrightarrow{I}\det(\vb I)\vb I\vb v = \vb v$ if $\vb v$ is an axial vector.

\subsection{Symmetry-constrained AHC}

Let us now consider the AHC $\sigma_{ij}$ defined by $j_i = \sigma_{ij} E_j$. To understand the magnetic-field dependence of the AHC once an external magnetic field is applied to the system, we may write the anomalous Hall current density as
\be
j_{i} = \sigma_{ij}(\vb B)E_j = \kappa_{ijk}E_j B_k,
\ee
where the structure of the tensor $\kappa_{ijk}$ fully reflects the symmetry of the unperturbed Hamiltonian (2) in the main text if the magnetic field is treated as a perturbation, which is valid in the weak-field limit.  Hence, $\kappa_{ijk}$ can be constrained by the generating symmetries $\qty{C_4,C_{2x},P}$. Under a rotation, the rank-3 tensor transform as
\be
\kappa_{ijk}\xrightarrow{R} \det(\vb R)^{N_\text{axial}}R_{il}R_{jm}R_{kn}\kappa_{lmn},
\ee
or in matrix notation,
\be
\boldsymbol{\kappa}_i\xrightarrow{R} \det(\vb R)^{N_\text{axial}} R_{il}\vb R\boldsymbol{\kappa}_{l}\vb R^T,
\ee
where $\boldsymbol{\kappa}_{l}$ is the $3\times 3$ matrix,
\bes
\boldsymbol{\kappa}_{l} = \mqty(\kappa_{lxx}&\kappa_{lxy}&\kappa_{lxz} \\ \kappa_{lyx}&\kappa_{lyy}&\kappa_{lyz} \\ \kappa_{lzx}&\kappa_{lzy}&\kappa_{lzz}),
\ees
and $N_\text{axial}$ is the number of indices in the tensor $\kappa$ that correspond to axial vectors. In this case, $N_\text{axial} = 1$. Let us first consider $C_{2x}$ symmetry. Under this symmetry transformation,
\bes
\begin{split}
    & \boldsymbol{\kappa}_x \xrightarrow{C_{2x}} \underbrace{R_{xx}}_{=1}\vb R_{2x}\boldsymbol{\kappa}_x\vb R_{2x}^T = \boldsymbol{\kappa}_x, \\
    & \boldsymbol{\kappa}_y \xrightarrow{C_{2x}} \underbrace{R_{yy}}_{=-1}\vb R_{2x}\boldsymbol{\kappa}_y\vb R_{2x}^T = \boldsymbol{\kappa}_y, \\
    & \boldsymbol{\kappa}_z \xrightarrow{C_{2x}} \underbrace{R_{zz}}_{=-1}\vb R_{2x}\boldsymbol{\kappa}_z\vb R_{2x}^T = \boldsymbol{\kappa}_z.
\end{split}
\ees
This translates to the following constraints on the tensor elements:
\bes
\begin{split}
    & \kappa_{xxy}= \kappa_{xxz} = \kappa_{xyx} = \kappa_{xzx} = 0,\\
    & \kappa_{yxx}= \kappa_{yyy} = \kappa_{yyz} = \kappa_{yzy} = \kappa_{yzz} = 0,\\
    & \kappa_{zxx}= \kappa_{zyy} = \kappa_{zyz} = \kappa_{zzy} = \kappa_{zzz} = 0.
\end{split}
\ees
Next, under $C_4$ transformation,
\bes
\begin{split}
    & \boldsymbol{\kappa}_x \xrightarrow{C_{4}} \underbrace{R_{xy}}_{=-1}\vb R_{4}\boldsymbol{\kappa}_y\vb R_{4}^T = \boldsymbol{\kappa}_x, \\
    & \boldsymbol{\kappa}_y \xrightarrow{C_{4}} \underbrace{R_{yx}}_{=1}\vb R_{4}\boldsymbol{\kappa}_x\vb R_{4}^T = \boldsymbol{\kappa}_y, \\
    & \boldsymbol{\kappa}_z \xrightarrow{C_{4}} \underbrace{R_{zz}}_{=1}\vb R_{4}\boldsymbol{\kappa}_z\vb R_{4}^T = \boldsymbol{\kappa}_z.
\end{split}
\ees
So $C_4$ symmetry imposes further constraints:
\bes
\begin{split}
    & \kappa_{xxx} = \kappa_{xyy} = \kappa_{xzz} = 0,\\
    & \kappa_{yxy} = \kappa_{yyx} = \kappa_{zxz} = \kappa_{zzx} = 0,\\
    & \kappa_{yxz} = -\kappa_{xyz},\quad \kappa_{yzx} = -\kappa_{xzy},\quad \kappa_{zyx} = -\kappa_{zxy}.
\end{split}
\ees
Inversion symmetry does not constrain $\kappa_{ijk}$ in this case. In summary, the anomalous Hall response matrices can be written as
\be
\boldsymbol{\kappa}_x = \mqty(0&0&0 \\ 0&0&\kappa_{xyz}\\ 0&\kappa_{xzy}&0),\quad \boldsymbol{\kappa}_y = \mqty(0&0&-\kappa_{xyz} \\ 0&0&0 \\ -\kappa_{xzy}&0&0),\quad \boldsymbol{\kappa}_z = \mqty(0&\kappa_{zxy}&0 \\ -\kappa_{zxy}&0&0 \\ 0&0&0).
\ee
However, as discussed in Sec.~III A in the main text, due to the topology of the TDSMs, $\kappa_{xzy} = \kappa_{zxy} = 0$. Therefore, the only nonzero tensor element corresponding to $\sigma_{xy}(\vb B)$ is $\kappa_{xyz}$, which agrees with the relation given by Eq.~(20) in the main text.

\subsection{Symmetry-constrained SHC}
The SHC $\sigma_{ab}^i$ is given by
\be
Q_a^i = \sigma_{ab}^i(\vb B) E_b =\chi^i_{abcd}E_bB_cB_d,
\ee
where the upper index $i$ denotes the direction of spin polarization. It is worth noting that the time-reversal symmetry dictates that the SHC must depend quadratically on the magnetic field to the lowest order. We can use the same symmetries $\qty{C_4,C_{2x},P}$ and techniques developed in the previous section to constrain the rank-5 tensor $\chi^i_{abcd}$. Under a rotation, the transformation rule can be written as
\be
\boldsymbol{\chi}^i_{ab}\xrightarrow{R} \det(\vb R)^{N_\text{axial}} R_{ij}R_{ak}R_{bl}\vb R\boldsymbol{\chi}^j_{kl}\vb R^T,
\ee
where $N_\text{axial}=3$ in this case. Let us again first focus on the constraints imposed by $C_{2x}$ on various $\boldsymbol{\chi}^i_{ab}$ matrices. After a straightforward but tedious calculation, we find
\bes
    \boldsymbol{\chi}^i_{ab} = \mqty(\chi^i_{abxx}&0&0 \\ 0&\chi^i_{abyy}&\chi^i_{abyz}\\ 0&\chi^i_{abzy}&\chi^i_{abzz})
\ees
where $i,a,b = \qty{xxx},\qty{xyy},\qty{xzz},\qty{xyz}$ (in any order), and in all other cases,
\bes
\boldsymbol{\chi}^i_{ab} = \mqty(0&\chi^i_{abxy}&\chi^i_{abxz} \\ \chi^i_{abyx}&0&0 \\ \chi^i_{abzx}&0&0).
\ees
An additional symmetry of the tensor elements is $\chi^i_{abij} = \chi^i_{abji}$. Next, we consider what constraints $C_4$ symmetry imposes on the tensor elements related to the SHC $\sigma_{xy}^i$. For example, for $\boldsymbol{\chi}^z_{xy}$, we need to impose the following constraints:
\bes
\begin{split}
   \boldsymbol{\chi}^z_{xy}\xrightarrow{C_4}\ &\underbrace{R_{zz}R_{xy}R_{yx}}_{=-1}\vb R_4\boldsymbol{\chi}^z_{yx}\vb R_4^T = \boldsymbol{\chi}^z_{xy}, \\
   \boldsymbol{\chi}^z_{yx}\xrightarrow{C_4}\ &\underbrace{R_{zz}R_{yx}R_{xy}}_{=-1}\vb R_4\boldsymbol{\chi}^z_{xy}\vb R_4^T = \boldsymbol{\chi}^z_{yx}.
  \end{split}
\ees
Solving the coupled matrix equations above leads to
\be
\boldsymbol{\chi}^z_{xy} = \mqty(\dmat[0]{\chi^z_{xyxx},\chi^z_{xyyy},\chi^z_{xyzz}}),\quad \boldsymbol{\chi}^z_{yx} = \mqty(\dmat[0]{-\chi^z_{xyyy},-\chi^z_{xyxx},-\chi^z_{xyzz}}).
\ee
Similarly, for the spin Hall tensors of other spin polarizations, we establish that 
\bes
\begin{split}
	\boldsymbol{\chi}^x_{xy}\xrightarrow{C_4}\ &R_{xy}R_{xy}R_{yx}\vb R_4\boldsymbol{\chi}^y_{yx}\vb R_4^T = \boldsymbol{\chi}^x_{xy},\\
	\boldsymbol{\chi}^y_{yx}\xrightarrow{C_4}\ &R_{yx}R_{yx}R_{xy}\vb R_4\boldsymbol{\chi}^x_{xy}\vb R_4^T = \boldsymbol{\chi}^y_{yx},\\
	\boldsymbol{\chi}^y_{xy}\xrightarrow{C_4}\ &R_{yx}R_{xy}R_{yx}\vb R_4\boldsymbol{\chi}^x_{yx}\vb R_4^T = \boldsymbol{\chi}^y_{xy},\\
	\boldsymbol{\chi}^x_{yx}\xrightarrow{C_4}\ &R_{xy}R_{yx}R_{xy}\vb R_4\boldsymbol{\chi}^y_{xy}\vb R_4^T = \boldsymbol{\chi}^x_{yx},
\end{split}
\ees
which leads to
\begin{align}
    &\boldsymbol{\chi}^x_{xy} = \mqty(0&0&\chi^x_{xyxz} \\ 0&0&0 \\\chi^x_{xyxz}&0&0),\quad &&\boldsymbol{\chi}^x_{yx} = \mqty(0&0&\chi^x_{yxxz} \\ 0&0&0 \\\chi^x_{yxxz}&0&0),\\
    &\boldsymbol{\chi}^y_{xy} = \mqty(0&0&0 \\ 0&0&-\chi^x_{yxxz} \\ 0&-\chi^x_{yxxz}&0),\quad && \boldsymbol{\chi}^y_{yx} = \mqty(0&0&0 \\ 0&0&-\chi^x_{xyxz} \\ 0&-\chi^x_{xyxz}&0).
\end{align}
Inversion symmetry again does  not place any constraint on the tensor $\chi^i_{abcd}$. Therefore, we have shown that the only nonzero tensor components associated with the SHC $\sigma^i_{xy}$ that are allowed by $D_{4h}$ symmetry are $\chi^x_{xyxz}$, $\chi^y_{xyyz}$, $\chi^z_{xyxx}$, $\chi^z_{xyyy}$, and $\chi^z_{xyzz}$. This indicates that the magnetic-field dependence of the SHC is given by Eq.~(18) in the main text. The other spin Hall tensor elements vanish due to the topological constraints in the TDSMs.


\bibliographystyle{apsrev4-1}
\bibliography{sc-TDSM_refs.bib}

\begin{thebibliography}{64}%
\makeatletter
\providecommand \@ifxundefined [1]{%
 \@ifx{#1\undefined}
}%
\providecommand \@ifnum [1]{%
 \ifnum #1\expandafter \@firstoftwo
 \else \expandafter \@secondoftwo
 \fi
}%
\providecommand \@ifx [1]{%
 \ifx #1\expandafter \@firstoftwo
 \else \expandafter \@secondoftwo
 \fi
}%
\providecommand \natexlab [1]{#1}%
\providecommand \enquote  [1]{``#1''}%
\providecommand \bibnamefont  [1]{#1}%
\providecommand \bibfnamefont [1]{#1}%
\providecommand \citenamefont [1]{#1}%
\providecommand \href@noop [0]{\@secondoftwo}%
\providecommand \href [0]{\begingroup \@sanitize@url \@href}%
\providecommand \@href[1]{\@@startlink{#1}\@@href}%
\providecommand \@@href[1]{\endgroup#1\@@endlink}%
\providecommand \@sanitize@url [0]{\catcode `\\12\catcode `\$12\catcode
  `\&12\catcode `\#12\catcode `\^12\catcode `\_12\catcode `\%12\relax}%
\providecommand \@@startlink[1]{}%
\providecommand \@@endlink[0]{}%
\providecommand \url  [0]{\begingroup\@sanitize@url \@url }%
\providecommand \@url [1]{\endgroup\@href {#1}{\urlprefix }}%
\providecommand \urlprefix  [0]{URL }%
\providecommand \Eprint [0]{\href }%
\providecommand \doibase [0]{http://dx.doi.org/}%
\providecommand \selectlanguage [0]{\@gobble}%
\providecommand \bibinfo  [0]{\@secondoftwo}%
\providecommand \bibfield  [0]{\@secondoftwo}%
\providecommand \translation [1]{[#1]}%
\providecommand \BibitemOpen [0]{}%
\providecommand \bibitemStop [0]{}%
\providecommand \bibitemNoStop [0]{.\EOS\space}%
\providecommand \EOS [0]{\spacefactor3000\relax}%
\providecommand \BibitemShut  [1]{\csname bibitem#1\endcsname}%
\let\auto@bib@innerbib\@empty
\bibitem [{\citenamefont {Dyakonov}\ and\ \citenamefont
  {Perel}(1971{\natexlab{a}})}]{DYAKONOV71PLA_spinHall}%
  \BibitemOpen
  \bibfield  {author} {\bibinfo {author} {\bibfnamefont {M.}~\bibnamefont
  {Dyakonov}}\ and\ \bibinfo {author} {\bibfnamefont {V.}~\bibnamefont
  {Perel}},\ }\href {\doibase https://doi.org/10.1016/0375-9601(71)90196-4}
  {\bibfield  {journal} {\bibinfo  {journal} {Phys. Lett. A}\ }\textbf
  {\bibinfo {volume} {35}},\ \bibinfo {pages} {459 } (\bibinfo {year}
  {1971}{\natexlab{a}})}\BibitemShut {NoStop}%
\bibitem [{\citenamefont {Dyakonov}\ and\ \citenamefont
  {Perel}(1971{\natexlab{b}})}]{Dyakonov71JETP_SHE}%
  \BibitemOpen
  \bibfield  {author} {\bibinfo {author} {\bibfnamefont {M.~I.}\ \bibnamefont
  {Dyakonov}}\ and\ \bibinfo {author} {\bibfnamefont {V.~I.}\ \bibnamefont
  {Perel}},\ }\href {\doibase
  https://ui.adsabs.harvard.edu/abs/1971JETPL..13..467D/abstract} {\bibfield
  {journal} {\bibinfo  {journal} {Sov. Phys. JETP}\ }\textbf {\bibinfo {volume}
  {13}},\ \bibinfo {pages} {467} (\bibinfo {year}
  {1971}{\natexlab{b}})}\BibitemShut {NoStop}%
\bibitem [{\citenamefont {Hirsch}(1999)}]{Hirsch99PRL_SHE}%
  \BibitemOpen
  \bibfield  {author} {\bibinfo {author} {\bibfnamefont {J.~E.}\ \bibnamefont
  {Hirsch}},\ }\href {\doibase 10.1103/PhysRevLett.83.1834} {\bibfield
  {journal} {\bibinfo  {journal} {Phys. Rev. Lett.}\ }\textbf {\bibinfo
  {volume} {83}},\ \bibinfo {pages} {1834} (\bibinfo {year}
  {1999})}\BibitemShut {NoStop}%
\bibitem [{\citenamefont {Zhang}(2000)}]{sZhang00PRL_SHE}%
  \BibitemOpen
  \bibfield  {author} {\bibinfo {author} {\bibfnamefont {S.}~\bibnamefont
  {Zhang}},\ }\href {\doibase 10.1103/PhysRevLett.85.393} {\bibfield  {journal}
  {\bibinfo  {journal} {Phys. Rev. Lett.}\ }\textbf {\bibinfo {volume} {85}},\
  \bibinfo {pages} {393} (\bibinfo {year} {2000})}\BibitemShut {NoStop}%
\bibitem [{\citenamefont {Sinova}\ \emph {et~al.}(2004)\citenamefont {Sinova},
  \citenamefont {Culcer}, \citenamefont {Niu}, \citenamefont {Sinitsyn},
  \citenamefont {Jungwirth},\ and\ \citenamefont {MacDonald}}]{Sinova04_SHE}%
  \BibitemOpen
  \bibfield  {author} {\bibinfo {author} {\bibfnamefont {J.}~\bibnamefont
  {Sinova}}, \bibinfo {author} {\bibfnamefont {D.}~\bibnamefont {Culcer}},
  \bibinfo {author} {\bibfnamefont {Q.}~\bibnamefont {Niu}}, \bibinfo {author}
  {\bibfnamefont {N.~A.}\ \bibnamefont {Sinitsyn}}, \bibinfo {author}
  {\bibfnamefont {T.}~\bibnamefont {Jungwirth}}, \ and\ \bibinfo {author}
  {\bibfnamefont {A.~H.}\ \bibnamefont {MacDonald}},\ }\href {\doibase
  10.1103/PhysRevLett.92.126603} {\bibfield  {journal} {\bibinfo  {journal}
  {Phys. Rev. Lett.}\ }\textbf {\bibinfo {volume} {92}},\ \bibinfo {pages}
  {126603} (\bibinfo {year} {2004})}\BibitemShut {NoStop}%
\bibitem [{\citenamefont {Vignale}(2010)}]{Vignale10_SHE}%
  \BibitemOpen
  \bibfield  {author} {\bibinfo {author} {\bibfnamefont {G.}~\bibnamefont
  {Vignale}},\ }\href@noop {} {\bibfield  {journal} {\bibinfo  {journal} {J.
  Supercond. Novel Magn.}\ }\textbf {\bibinfo {volume} {23}},\ \bibinfo {pages}
  {3} (\bibinfo {year} {2010})}\BibitemShut {NoStop}%
\bibitem [{\citenamefont {Hoffmann}(2013)}]{aHoffmann13ieee_SHE}%
  \BibitemOpen
  \bibfield  {author} {\bibinfo {author} {\bibfnamefont {A.}~\bibnamefont
  {Hoffmann}},\ }\href {\doibase 10.1109/TMAG.2013.2262947} {\bibfield
  {journal} {\bibinfo  {journal} {IEEE Trans. Magn.}\ }\textbf {\bibinfo
  {volume} {49}},\ \bibinfo {pages} {5172} (\bibinfo {year}
  {2013})}\BibitemShut {NoStop}%
\bibitem [{\citenamefont {Sinova}\ \emph {et~al.}(2015)\citenamefont {Sinova},
  \citenamefont {Valenzuela}, \citenamefont {Wunderlich}, \citenamefont
  {Back},\ and\ \citenamefont {Jungwirth}}]{Sinova15RMP_SHE}%
  \BibitemOpen
  \bibfield  {author} {\bibinfo {author} {\bibfnamefont {J.}~\bibnamefont
  {Sinova}}, \bibinfo {author} {\bibfnamefont {S.~O.}\ \bibnamefont
  {Valenzuela}}, \bibinfo {author} {\bibfnamefont {J.}~\bibnamefont
  {Wunderlich}}, \bibinfo {author} {\bibfnamefont {C.~H.}\ \bibnamefont
  {Back}}, \ and\ \bibinfo {author} {\bibfnamefont {T.}~\bibnamefont
  {Jungwirth}},\ }\href {\doibase 10.1103/RevModPhys.87.1213} {\bibfield
  {journal} {\bibinfo  {journal} {Rev. Mod. Phys.}\ }\textbf {\bibinfo {volume}
  {87}},\ \bibinfo {pages} {1213} (\bibinfo {year} {2015})}\BibitemShut
  {NoStop}%
\bibitem [{\citenamefont {Ando}\ \emph {et~al.}(2011)\citenamefont {Ando},
  \citenamefont {Takahashi}, \citenamefont {Ieda}, \citenamefont {Kajiwara},
  \citenamefont {Nakayama}, \citenamefont {Yoshino}, \citenamefont {Harii},
  \citenamefont {Fujikawa}, \citenamefont {Matsuo}, \citenamefont {Maekawa},\
  and\ \citenamefont {Saitoh}}]{Saitoh11JAP_ISHE}%
  \BibitemOpen
  \bibfield  {author} {\bibinfo {author} {\bibfnamefont {K.}~\bibnamefont
  {Ando}}, \bibinfo {author} {\bibfnamefont {S.}~\bibnamefont {Takahashi}},
  \bibinfo {author} {\bibfnamefont {J.}~\bibnamefont {Ieda}}, \bibinfo {author}
  {\bibfnamefont {Y.}~\bibnamefont {Kajiwara}}, \bibinfo {author}
  {\bibfnamefont {H.}~\bibnamefont {Nakayama}}, \bibinfo {author}
  {\bibfnamefont {T.}~\bibnamefont {Yoshino}}, \bibinfo {author} {\bibfnamefont
  {K.}~\bibnamefont {Harii}}, \bibinfo {author} {\bibfnamefont
  {Y.}~\bibnamefont {Fujikawa}}, \bibinfo {author} {\bibfnamefont
  {M.}~\bibnamefont {Matsuo}}, \bibinfo {author} {\bibfnamefont
  {S.}~\bibnamefont {Maekawa}}, \ and\ \bibinfo {author} {\bibfnamefont
  {E.}~\bibnamefont {Saitoh}},\ }\href {\doibase 10.1063/1.3587173} {\bibfield
  {journal} {\bibinfo  {journal} {J. Appl. Phys.}\ }\textbf {\bibinfo {volume}
  {109}},\ \bibinfo {pages} {103913} (\bibinfo {year} {2011})}\BibitemShut
  {NoStop}%
\bibitem [{\citenamefont {S{\'a}nchez}\ \emph {et~al.}(2013)\citenamefont
  {S{\'a}nchez}, \citenamefont {Vila}, \citenamefont {Desfonds}, \citenamefont
  {Gambarelli}, \citenamefont {Attan{\'e}}, \citenamefont {De~Teresa},
  \citenamefont {Mag{\'e}n},\ and\ \citenamefont
  {Fert}}]{Sanchez13NatComm_SC-conv}%
  \BibitemOpen
  \bibfield  {author} {\bibinfo {author} {\bibfnamefont {J.~C.~R.}\
  \bibnamefont {S{\'a}nchez}}, \bibinfo {author} {\bibfnamefont
  {L.}~\bibnamefont {Vila}}, \bibinfo {author} {\bibfnamefont {G.}~\bibnamefont
  {Desfonds}}, \bibinfo {author} {\bibfnamefont {S.}~\bibnamefont
  {Gambarelli}}, \bibinfo {author} {\bibfnamefont {J.~P.}\ \bibnamefont
  {Attan{\'e}}}, \bibinfo {author} {\bibfnamefont {J.~M.}\ \bibnamefont
  {De~Teresa}}, \bibinfo {author} {\bibfnamefont {C.}~\bibnamefont
  {Mag{\'e}n}}, \ and\ \bibinfo {author} {\bibfnamefont {A.}~\bibnamefont
  {Fert}},\ }\href {https://doi.org/10.1038/ncomms3944} {\bibfield  {journal}
  {\bibinfo  {journal} {Nat. Commun.}\ }\textbf {\bibinfo {volume} {4}},\
  \bibinfo {pages} {2944} (\bibinfo {year} {2013})}\BibitemShut {NoStop}%
\bibitem [{\citenamefont {Shiomi}\ \emph {et~al.}(2014)\citenamefont {Shiomi},
  \citenamefont {Nomura}, \citenamefont {Kajiwara}, \citenamefont {Eto},
  \citenamefont {Novak}, \citenamefont {Segawa}, \citenamefont {Ando},\ and\
  \citenamefont {Saitoh}}]{Saitoh14PRL_SC-TI}%
  \BibitemOpen
  \bibfield  {author} {\bibinfo {author} {\bibfnamefont {Y.}~\bibnamefont
  {Shiomi}}, \bibinfo {author} {\bibfnamefont {K.}~\bibnamefont {Nomura}},
  \bibinfo {author} {\bibfnamefont {Y.}~\bibnamefont {Kajiwara}}, \bibinfo
  {author} {\bibfnamefont {K.}~\bibnamefont {Eto}}, \bibinfo {author}
  {\bibfnamefont {M.}~\bibnamefont {Novak}}, \bibinfo {author} {\bibfnamefont
  {K.}~\bibnamefont {Segawa}}, \bibinfo {author} {\bibfnamefont
  {Y.}~\bibnamefont {Ando}}, \ and\ \bibinfo {author} {\bibfnamefont
  {E.}~\bibnamefont {Saitoh}},\ }\href {\doibase
  10.1103/PhysRevLett.113.196601} {\bibfield  {journal} {\bibinfo  {journal}
  {Phys. Rev. Lett.}\ }\textbf {\bibinfo {volume} {113}},\ \bibinfo {pages}
  {196601} (\bibinfo {year} {2014})}\BibitemShut {NoStop}%
\bibitem [{\citenamefont {Rojas-S\'anchez}\ \emph {et~al.}(2016)\citenamefont
  {Rojas-S\'anchez}, \citenamefont {Oyarz\'un}, \citenamefont {Fu},
  \citenamefont {Marty}, \citenamefont {Vergnaud}, \citenamefont {Gambarelli},
  \citenamefont {Vila}, \citenamefont {Jamet}, \citenamefont {Ohtsubo},
  \citenamefont {Taleb-Ibrahimi}, \citenamefont {Le~F\`evre}, \citenamefont
  {Bertran}, \citenamefont {Reyren}, \citenamefont {George},\ and\
  \citenamefont {Fert}}]{Sanchez16PRL_SCC-a-Sn}%
  \BibitemOpen
  \bibfield  {author} {\bibinfo {author} {\bibfnamefont {J.-C.}\ \bibnamefont
  {Rojas-S\'anchez}}, \bibinfo {author} {\bibfnamefont {S.}~\bibnamefont
  {Oyarz\'un}}, \bibinfo {author} {\bibfnamefont {Y.}~\bibnamefont {Fu}},
  \bibinfo {author} {\bibfnamefont {A.}~\bibnamefont {Marty}}, \bibinfo
  {author} {\bibfnamefont {C.}~\bibnamefont {Vergnaud}}, \bibinfo {author}
  {\bibfnamefont {S.}~\bibnamefont {Gambarelli}}, \bibinfo {author}
  {\bibfnamefont {L.}~\bibnamefont {Vila}}, \bibinfo {author} {\bibfnamefont
  {M.}~\bibnamefont {Jamet}}, \bibinfo {author} {\bibfnamefont
  {Y.}~\bibnamefont {Ohtsubo}}, \bibinfo {author} {\bibfnamefont
  {A.}~\bibnamefont {Taleb-Ibrahimi}}, \bibinfo {author} {\bibfnamefont
  {P.}~\bibnamefont {Le~F\`evre}}, \bibinfo {author} {\bibfnamefont
  {F.}~\bibnamefont {Bertran}}, \bibinfo {author} {\bibfnamefont
  {N.}~\bibnamefont {Reyren}}, \bibinfo {author} {\bibfnamefont {J.-M.}\
  \bibnamefont {George}}, \ and\ \bibinfo {author} {\bibfnamefont
  {A.}~\bibnamefont {Fert}},\ }\href {\doibase 10.1103/PhysRevLett.116.096602}
  {\bibfield  {journal} {\bibinfo  {journal} {Phys. Rev. Lett.}\ }\textbf
  {\bibinfo {volume} {116}},\ \bibinfo {pages} {096602} (\bibinfo {year}
  {2016})}\BibitemShut {NoStop}%
\bibitem [{\citenamefont {Sun}\ \emph {et~al.}(2016)\citenamefont {Sun},
  \citenamefont {Zhang}, \citenamefont {Felser},\ and\ \citenamefont
  {Yan}}]{ySun16PRL_SHE-WSM}%
  \BibitemOpen
  \bibfield  {author} {\bibinfo {author} {\bibfnamefont {Y.}~\bibnamefont
  {Sun}}, \bibinfo {author} {\bibfnamefont {Y.}~\bibnamefont {Zhang}}, \bibinfo
  {author} {\bibfnamefont {C.}~\bibnamefont {Felser}}, \ and\ \bibinfo {author}
  {\bibfnamefont {B.}~\bibnamefont {Yan}},\ }\href {\doibase
  10.1103/PhysRevLett.117.146403} {\bibfield  {journal} {\bibinfo  {journal}
  {Phys. Rev. Lett.}\ }\textbf {\bibinfo {volume} {117}},\ \bibinfo {pages}
  {146403} (\bibinfo {year} {2016})}\BibitemShut {NoStop}%
\bibitem [{\citenamefont {Sun}\ \emph {et~al.}(2017)\citenamefont {Sun},
  \citenamefont {Zhang}, \citenamefont {Liu}, \citenamefont {Felser},\ and\
  \citenamefont {Yan}}]{ySun17PRB_SHE-DNL}%
  \BibitemOpen
  \bibfield  {author} {\bibinfo {author} {\bibfnamefont {Y.}~\bibnamefont
  {Sun}}, \bibinfo {author} {\bibfnamefont {Y.}~\bibnamefont {Zhang}}, \bibinfo
  {author} {\bibfnamefont {C.-X.}\ \bibnamefont {Liu}}, \bibinfo {author}
  {\bibfnamefont {C.}~\bibnamefont {Felser}}, \ and\ \bibinfo {author}
  {\bibfnamefont {B.}~\bibnamefont {Yan}},\ }\href {\doibase
  10.1103/PhysRevB.95.235104} {\bibfield  {journal} {\bibinfo  {journal} {Phys.
  Rev. B}\ }\textbf {\bibinfo {volume} {95}},\ \bibinfo {pages} {235104}
  (\bibinfo {year} {2017})}\BibitemShut {NoStop}%
\bibitem [{\citenamefont {Han}\ \emph {et~al.}(2018)\citenamefont {Han},
  \citenamefont {Otani},\ and\ \citenamefont
  {Maekawa}}]{Han18NPJ-QM_SC-review}%
  \BibitemOpen
  \bibfield  {author} {\bibinfo {author} {\bibfnamefont {W.}~\bibnamefont
  {Han}}, \bibinfo {author} {\bibfnamefont {Y.}~\bibnamefont {Otani}}, \ and\
  \bibinfo {author} {\bibfnamefont {S.}~\bibnamefont {Maekawa}},\ }\href
  {\doibase 10.1038/s41535-018-0100-9} {\bibfield  {journal} {\bibinfo
  {journal} {npj Quantum Materials}\ }\textbf {\bibinfo {volume} {3}},\
  \bibinfo {pages} {27} (\bibinfo {year} {2018})}\BibitemShut {NoStop}%
\bibitem [{\citenamefont {Derunova}\ \emph {et~al.}(2019)\citenamefont
  {Derunova}, \citenamefont {Sun}, \citenamefont {Felser}, \citenamefont
  {Parkin}, \citenamefont {Yan},\ and\ \citenamefont
  {Ali}}]{Derunova19_ISHE-SC}%
  \BibitemOpen
  \bibfield  {author} {\bibinfo {author} {\bibfnamefont {E.}~\bibnamefont
  {Derunova}}, \bibinfo {author} {\bibfnamefont {Y.}~\bibnamefont {Sun}},
  \bibinfo {author} {\bibfnamefont {C.}~\bibnamefont {Felser}}, \bibinfo
  {author} {\bibfnamefont {S.~S.~P.}\ \bibnamefont {Parkin}}, \bibinfo {author}
  {\bibfnamefont {B.}~\bibnamefont {Yan}}, \ and\ \bibinfo {author}
  {\bibfnamefont {M.~N.}\ \bibnamefont {Ali}},\ }\href
  {https://advances.sciencemag.org/content/5/4/eaav8575} {\bibfield  {journal}
  {\bibinfo  {journal} {Sci. Adv.}\ }\textbf {\bibinfo {volume} {5}} (\bibinfo
  {year} {2019})}\BibitemShut {NoStop}%
\bibitem [{\citenamefont {Vaz}\ \emph {et~al.}(2019)\citenamefont {Vaz},
  \citenamefont {No{\"e}l}, \citenamefont {Johansson}, \citenamefont
  {G{\"o}bel}, \citenamefont {Bruno}, \citenamefont {Singh}, \citenamefont
  {McKeown-Walker}, \citenamefont {Trier}, \citenamefont {Vicente-Arche},
  \citenamefont {Sander}, \citenamefont {Valencia}, \citenamefont {Bruneel},
  \citenamefont {Vivek}, \citenamefont {Gabay}, \citenamefont {Bergeal},
  \citenamefont {Baumberger}, \citenamefont {Okuno}, \citenamefont
  {Barth{\'e}l{\'e}my}, \citenamefont {Fert}, \citenamefont {Vila},
  \citenamefont {Mertig}, \citenamefont {Attan{\'e}},\ and\ \citenamefont
  {Bibes}}]{Vaz19NM_sc-STO3}%
  \BibitemOpen
  \bibfield  {author} {\bibinfo {author} {\bibfnamefont {D.~C.}\ \bibnamefont
  {Vaz}}, \bibinfo {author} {\bibfnamefont {P.}~\bibnamefont {No{\"e}l}},
  \bibinfo {author} {\bibfnamefont {A.}~\bibnamefont {Johansson}}, \bibinfo
  {author} {\bibfnamefont {B.}~\bibnamefont {G{\"o}bel}}, \bibinfo {author}
  {\bibfnamefont {F.~Y.}\ \bibnamefont {Bruno}}, \bibinfo {author}
  {\bibfnamefont {G.}~\bibnamefont {Singh}}, \bibinfo {author} {\bibfnamefont
  {S.}~\bibnamefont {McKeown-Walker}}, \bibinfo {author} {\bibfnamefont
  {F.}~\bibnamefont {Trier}}, \bibinfo {author} {\bibfnamefont {L.~M.}\
  \bibnamefont {Vicente-Arche}}, \bibinfo {author} {\bibfnamefont
  {A.}~\bibnamefont {Sander}}, \bibinfo {author} {\bibfnamefont
  {S.}~\bibnamefont {Valencia}}, \bibinfo {author} {\bibfnamefont
  {P.}~\bibnamefont {Bruneel}}, \bibinfo {author} {\bibfnamefont
  {M.}~\bibnamefont {Vivek}}, \bibinfo {author} {\bibfnamefont
  {M.}~\bibnamefont {Gabay}}, \bibinfo {author} {\bibfnamefont
  {N.}~\bibnamefont {Bergeal}}, \bibinfo {author} {\bibfnamefont
  {F.}~\bibnamefont {Baumberger}}, \bibinfo {author} {\bibfnamefont
  {H.}~\bibnamefont {Okuno}}, \bibinfo {author} {\bibfnamefont
  {A.}~\bibnamefont {Barth{\'e}l{\'e}my}}, \bibinfo {author} {\bibfnamefont
  {A.}~\bibnamefont {Fert}}, \bibinfo {author} {\bibfnamefont {L.}~\bibnamefont
  {Vila}}, \bibinfo {author} {\bibfnamefont {I.}~\bibnamefont {Mertig}},
  \bibinfo {author} {\bibfnamefont {J.-P.}\ \bibnamefont {Attan{\'e}}}, \ and\
  \bibinfo {author} {\bibfnamefont {M.}~\bibnamefont {Bibes}},\ }\href
  {\doibase 10.1038/s41563-019-0467-4} {\bibfield  {journal} {\bibinfo
  {journal} {Nat. Mater.}\ }\textbf {\bibinfo {volume} {18}},\ \bibinfo {pages}
  {1187} (\bibinfo {year} {2019})}\BibitemShut {NoStop}%
\bibitem [{\citenamefont {Zhang}\ \emph
  {et~al.}(2019{\natexlab{a}})\citenamefont {Zhang}, \citenamefont {Burkov},
  \citenamefont {Martin},\ and\ \citenamefont
  {Heinonen}}]{sZhang19PRL_sc-mWSM}%
  \BibitemOpen
  \bibfield  {author} {\bibinfo {author} {\bibfnamefont {S.~S.-L.}\
  \bibnamefont {Zhang}}, \bibinfo {author} {\bibfnamefont {A.~A.}\ \bibnamefont
  {Burkov}}, \bibinfo {author} {\bibfnamefont {I.}~\bibnamefont {Martin}}, \
  and\ \bibinfo {author} {\bibfnamefont {O.~G.}\ \bibnamefont {Heinonen}},\
  }\href {\doibase 10.1103/PhysRevLett.123.187201} {\bibfield  {journal}
  {\bibinfo  {journal} {Phys. Rev. Lett.}\ }\textbf {\bibinfo {volume} {123}},\
  \bibinfo {pages} {187201} (\bibinfo {year} {2019}{\natexlab{a}})}\BibitemShut
  {NoStop}%
\bibitem [{\citenamefont {Safi}\ \emph {et~al.}(2020)\citenamefont {Safi},
  \citenamefont {Zhang}, \citenamefont {Fan}, \citenamefont {Guo},
  \citenamefont {Han}, \citenamefont {Rosenberg}, \citenamefont {Ross},
  \citenamefont {Tserkovnyak},\ and\ \citenamefont
  {Liu}}]{luqiao20NC_sc-VO2phase}%
  \BibitemOpen
  \bibfield  {author} {\bibinfo {author} {\bibfnamefont {T.~S.}\ \bibnamefont
  {Safi}}, \bibinfo {author} {\bibfnamefont {P.}~\bibnamefont {Zhang}},
  \bibinfo {author} {\bibfnamefont {Y.}~\bibnamefont {Fan}}, \bibinfo {author}
  {\bibfnamefont {Z.}~\bibnamefont {Guo}}, \bibinfo {author} {\bibfnamefont
  {J.}~\bibnamefont {Han}}, \bibinfo {author} {\bibfnamefont {E.~R.}\
  \bibnamefont {Rosenberg}}, \bibinfo {author} {\bibfnamefont {C.}~\bibnamefont
  {Ross}}, \bibinfo {author} {\bibfnamefont {Y.}~\bibnamefont {Tserkovnyak}}, \
  and\ \bibinfo {author} {\bibfnamefont {L.}~\bibnamefont {Liu}},\ }\href
  {\doibase 10.1038/s41467-020-14388-9} {\bibfield  {journal} {\bibinfo
  {journal} {Nat. Commun.}\ }\textbf {\bibinfo {volume} {11}},\ \bibinfo
  {pages} {476} (\bibinfo {year} {2020})}\BibitemShut {NoStop}%
\bibitem [{\citenamefont {Yen}\ and\ \citenamefont
  {Guo}(2020)}]{Yen&Guo20PRB_SHE_SNE-DSM}%
  \BibitemOpen
  \bibfield  {author} {\bibinfo {author} {\bibfnamefont {Y.}~\bibnamefont
  {Yen}}\ and\ \bibinfo {author} {\bibfnamefont {G.-Y.}\ \bibnamefont {Guo}},\
  }\href {\doibase 10.1103/PhysRevB.101.064430} {\bibfield  {journal} {\bibinfo
   {journal} {Phys. Rev. B}\ }\textbf {\bibinfo {volume} {101}},\ \bibinfo
  {pages} {064430} (\bibinfo {year} {2020})}\BibitemShut {NoStop}%
\bibitem [{\citenamefont {Taguchi}\ \emph {et~al.}(2020)\citenamefont
  {Taguchi}, \citenamefont {Oshima}, \citenamefont {Yamaguchi}, \citenamefont
  {Hashimoto}, \citenamefont {Tanaka},\ and\ \citenamefont
  {Sato}}]{Katsuhsia20PRB_DSM-SHE}%
  \BibitemOpen
  \bibfield  {author} {\bibinfo {author} {\bibfnamefont {K.}~\bibnamefont
  {Taguchi}}, \bibinfo {author} {\bibfnamefont {D.}~\bibnamefont {Oshima}},
  \bibinfo {author} {\bibfnamefont {Y.}~\bibnamefont {Yamaguchi}}, \bibinfo
  {author} {\bibfnamefont {T.}~\bibnamefont {Hashimoto}}, \bibinfo {author}
  {\bibfnamefont {Y.}~\bibnamefont {Tanaka}}, \ and\ \bibinfo {author}
  {\bibfnamefont {M.}~\bibnamefont {Sato}},\ }\href {\doibase
  10.1103/PhysRevB.101.235201} {\bibfield  {journal} {\bibinfo  {journal}
  {Phys. Rev. B}\ }\textbf {\bibinfo {volume} {101}},\ \bibinfo {pages}
  {235201} (\bibinfo {year} {2020})}\BibitemShut {NoStop}%
\bibitem [{\citenamefont {Farzaneh}\ and\ \citenamefont
  {Rakheja}(2020)}]{Farzaneh20PRM_ISHE-TI}%
  \BibitemOpen
  \bibfield  {author} {\bibinfo {author} {\bibfnamefont {S.~M.}\ \bibnamefont
  {Farzaneh}}\ and\ \bibinfo {author} {\bibfnamefont {S.}~\bibnamefont
  {Rakheja}},\ }\href {\doibase 10.1103/PhysRevMaterials.4.114202} {\bibfield
  {journal} {\bibinfo  {journal} {Phys. Rev. Materials}\ }\textbf {\bibinfo
  {volume} {4}},\ \bibinfo {pages} {114202} (\bibinfo {year}
  {2020})}\BibitemShut {NoStop}%
\bibitem [{\citenamefont {Zhang}\ \emph {et~al.}(2014)\citenamefont {Zhang},
  \citenamefont {Chen},\ and\ \citenamefont {Zhang}}]{slZhang14EPL}%
  \BibitemOpen
  \bibfield  {author} {\bibinfo {author} {\bibfnamefont {S.~S.-L.}\
  \bibnamefont {Zhang}}, \bibinfo {author} {\bibfnamefont {K.}~\bibnamefont
  {Chen}}, \ and\ \bibinfo {author} {\bibfnamefont {S.}~\bibnamefont {Zhang}},\
  }\href {http://stacks.iop.org/0295-5075/106/i=6/a=67007} {\bibfield
  {journal} {\bibinfo  {journal} {Europhys. Lett.}\ }\textbf {\bibinfo {volume}
  {106}},\ \bibinfo {pages} {67007} (\bibinfo {year} {2014})}\BibitemShut
  {NoStop}%
\bibitem [{\citenamefont {MacNeill}\ \emph {et~al.}(2017)\citenamefont
  {MacNeill}, \citenamefont {Stiehl}, \citenamefont {Guimaraes}, \citenamefont
  {Buhrman}, \citenamefont {Park},\ and\ \citenamefont
  {Ralph}}]{MacNeill17NP_SOT-sym-WTe2}%
  \BibitemOpen
  \bibfield  {author} {\bibinfo {author} {\bibfnamefont {D.}~\bibnamefont
  {MacNeill}}, \bibinfo {author} {\bibfnamefont {G.~M.}\ \bibnamefont
  {Stiehl}}, \bibinfo {author} {\bibfnamefont {M.~H.~D.}\ \bibnamefont
  {Guimaraes}}, \bibinfo {author} {\bibfnamefont {R.~A.}\ \bibnamefont
  {Buhrman}}, \bibinfo {author} {\bibfnamefont {J.}~\bibnamefont {Park}}, \
  and\ \bibinfo {author} {\bibfnamefont {D.~C.}\ \bibnamefont {Ralph}},\ }\href
  {\doibase 10.1038/nphys3933} {\bibfield  {journal} {\bibinfo  {journal} {Nat.
  Phys.}\ }\textbf {\bibinfo {volume} {13}},\ \bibinfo {pages} {300} (\bibinfo
  {year} {2017})}\BibitemShut {NoStop}%
\bibitem [{\citenamefont {Xue}\ and\ \citenamefont
  {Haney}(2020)}]{Xue20PRB_staggered_SHE}%
  \BibitemOpen
  \bibfield  {author} {\bibinfo {author} {\bibfnamefont {F.}~\bibnamefont
  {Xue}}\ and\ \bibinfo {author} {\bibfnamefont {P.~M.}\ \bibnamefont
  {Haney}},\ }\href {\doibase 10.1103/PhysRevB.102.195146} {\bibfield
  {journal} {\bibinfo  {journal} {Phys. Rev. B}\ }\textbf {\bibinfo {volume}
  {102}},\ \bibinfo {pages} {195146} (\bibinfo {year} {2020})}\BibitemShut
  {NoStop}%
\bibitem [{\citenamefont {Zhao}\ \emph {et~al.}(2020)\citenamefont {Zhao},
  \citenamefont {Khokhriakov}, \citenamefont {Zhang}, \citenamefont {Fu},
  \citenamefont {Karpiak}, \citenamefont {Hoque}, \citenamefont {Xu},
  \citenamefont {Jiang}, \citenamefont {Yan},\ and\ \citenamefont
  {Dash}}]{Zhao20PRR_CSC_WTe2}%
  \BibitemOpen
  \bibfield  {author} {\bibinfo {author} {\bibfnamefont {B.}~\bibnamefont
  {Zhao}}, \bibinfo {author} {\bibfnamefont {D.}~\bibnamefont {Khokhriakov}},
  \bibinfo {author} {\bibfnamefont {Y.}~\bibnamefont {Zhang}}, \bibinfo
  {author} {\bibfnamefont {H.}~\bibnamefont {Fu}}, \bibinfo {author}
  {\bibfnamefont {B.}~\bibnamefont {Karpiak}}, \bibinfo {author} {\bibfnamefont
  {A.~M.}\ \bibnamefont {Hoque}}, \bibinfo {author} {\bibfnamefont
  {X.}~\bibnamefont {Xu}}, \bibinfo {author} {\bibfnamefont {Y.}~\bibnamefont
  {Jiang}}, \bibinfo {author} {\bibfnamefont {B.}~\bibnamefont {Yan}}, \ and\
  \bibinfo {author} {\bibfnamefont {S.~P.}\ \bibnamefont {Dash}},\ }\href
  {\doibase 10.1103/PhysRevResearch.2.013286} {\bibfield  {journal} {\bibinfo
  {journal} {Phys. Rev. Research}\ }\textbf {\bibinfo {volume} {2}},\ \bibinfo
  {pages} {013286} (\bibinfo {year} {2020})}\BibitemShut {NoStop}%
\bibitem [{\citenamefont {Song}\ \emph {et~al.}(2020)\citenamefont {Song},
  \citenamefont {Hsu}, \citenamefont {Vignale}, \citenamefont {Zhao},
  \citenamefont {Liu}, \citenamefont {Deng}, \citenamefont {Fu}, \citenamefont
  {Liu}, \citenamefont {Zhang}, \citenamefont {Lin}, \citenamefont {Pereira},\
  and\ \citenamefont {Loh}}]{Song20nmat_MoTe2_SHE}%
  \BibitemOpen
  \bibfield  {author} {\bibinfo {author} {\bibfnamefont {P.}~\bibnamefont
  {Song}}, \bibinfo {author} {\bibfnamefont {C.-H.}\ \bibnamefont {Hsu}},
  \bibinfo {author} {\bibfnamefont {G.}~\bibnamefont {Vignale}}, \bibinfo
  {author} {\bibfnamefont {M.}~\bibnamefont {Zhao}}, \bibinfo {author}
  {\bibfnamefont {J.}~\bibnamefont {Liu}}, \bibinfo {author} {\bibfnamefont
  {Y.}~\bibnamefont {Deng}}, \bibinfo {author} {\bibfnamefont {W.}~\bibnamefont
  {Fu}}, \bibinfo {author} {\bibfnamefont {Y.}~\bibnamefont {Liu}}, \bibinfo
  {author} {\bibfnamefont {Y.}~\bibnamefont {Zhang}}, \bibinfo {author}
  {\bibfnamefont {H.}~\bibnamefont {Lin}}, \bibinfo {author} {\bibfnamefont
  {V.~M.}\ \bibnamefont {Pereira}}, \ and\ \bibinfo {author} {\bibfnamefont
  {K.~P.}\ \bibnamefont {Loh}},\ }\href {\doibase 10.1038/s41563-019-0600-4}
  {\bibfield  {journal} {\bibinfo  {journal} {Nat. Mater.}\ }\textbf {\bibinfo
  {volume} {19}},\ \bibinfo {pages} {292} (\bibinfo {year} {2020})}\BibitemShut
  {NoStop}%
\bibitem [{\citenamefont {Amin}\ \emph {et~al.}(2019)\citenamefont {Amin},
  \citenamefont {Li}, \citenamefont {Stiles},\ and\ \citenamefont
  {Haney}}]{Amin19PRB_SHE_FM}%
  \BibitemOpen
  \bibfield  {author} {\bibinfo {author} {\bibfnamefont {V.~P.}\ \bibnamefont
  {Amin}}, \bibinfo {author} {\bibfnamefont {J.}~\bibnamefont {Li}}, \bibinfo
  {author} {\bibfnamefont {M.~D.}\ \bibnamefont {Stiles}}, \ and\ \bibinfo
  {author} {\bibfnamefont {P.~M.}\ \bibnamefont {Haney}},\ }\href {\doibase
  10.1103/PhysRevB.99.220405} {\bibfield  {journal} {\bibinfo  {journal} {Phys.
  Rev. B}\ }\textbf {\bibinfo {volume} {99}},\ \bibinfo {pages} {220405}
  (\bibinfo {year} {2019})}\BibitemShut {NoStop}%
\bibitem [{\citenamefont {Davidson}\ \emph {et~al.}(2020)\citenamefont
  {Davidson}, \citenamefont {Amin}, \citenamefont {Aljuaid}, \citenamefont
  {Haney},\ and\ \citenamefont {Fan}}]{Davidson20PLA_FMspin}%
  \BibitemOpen
  \bibfield  {author} {\bibinfo {author} {\bibfnamefont {A.}~\bibnamefont
  {Davidson}}, \bibinfo {author} {\bibfnamefont {V.~P.}\ \bibnamefont {Amin}},
  \bibinfo {author} {\bibfnamefont {W.~S.}\ \bibnamefont {Aljuaid}}, \bibinfo
  {author} {\bibfnamefont {P.~M.}\ \bibnamefont {Haney}}, \ and\ \bibinfo
  {author} {\bibfnamefont {X.}~\bibnamefont {Fan}},\ }\href {\doibase
  10.1016/j.physleta.2019.126228} {\bibfield  {journal} {\bibinfo  {journal}
  {Phys. Lett. A}\ }\textbf {\bibinfo {volume} {384}},\ \bibinfo {pages}
  {126228} (\bibinfo {year} {2020})}\BibitemShut {NoStop}%
\bibitem [{\citenamefont {Wang}\ \emph {et~al.}(2012)\citenamefont {Wang},
  \citenamefont {Sun}, \citenamefont {Chen}, \citenamefont {Franchini},
  \citenamefont {Xu}, \citenamefont {Weng}, \citenamefont {Dai},\ and\
  \citenamefont {Fang}}]{zWang12PRB_Na3Bi}%
  \BibitemOpen
  \bibfield  {author} {\bibinfo {author} {\bibfnamefont {Z.}~\bibnamefont
  {Wang}}, \bibinfo {author} {\bibfnamefont {Y.}~\bibnamefont {Sun}}, \bibinfo
  {author} {\bibfnamefont {X.-Q.}\ \bibnamefont {Chen}}, \bibinfo {author}
  {\bibfnamefont {C.}~\bibnamefont {Franchini}}, \bibinfo {author}
  {\bibfnamefont {G.}~\bibnamefont {Xu}}, \bibinfo {author} {\bibfnamefont
  {H.}~\bibnamefont {Weng}}, \bibinfo {author} {\bibfnamefont {X.}~\bibnamefont
  {Dai}}, \ and\ \bibinfo {author} {\bibfnamefont {Z.}~\bibnamefont {Fang}},\
  }\href {\doibase 10.1103/PhysRevB.85.195320} {\bibfield  {journal} {\bibinfo
  {journal} {Phys. Rev. B}\ }\textbf {\bibinfo {volume} {85}},\ \bibinfo
  {pages} {195320} (\bibinfo {year} {2012})}\BibitemShut {NoStop}%
\bibitem [{\citenamefont {Wang}\ \emph {et~al.}(2013)\citenamefont {Wang},
  \citenamefont {Weng}, \citenamefont {Wu}, \citenamefont {Dai},\ and\
  \citenamefont {Fang}}]{zWang13prb_Cd3As2}%
  \BibitemOpen
  \bibfield  {author} {\bibinfo {author} {\bibfnamefont {Z.}~\bibnamefont
  {Wang}}, \bibinfo {author} {\bibfnamefont {H.}~\bibnamefont {Weng}}, \bibinfo
  {author} {\bibfnamefont {Q.}~\bibnamefont {Wu}}, \bibinfo {author}
  {\bibfnamefont {X.}~\bibnamefont {Dai}}, \ and\ \bibinfo {author}
  {\bibfnamefont {Z.}~\bibnamefont {Fang}},\ }\href {\doibase
  10.1103/PhysRevB.88.125427} {\bibfield  {journal} {\bibinfo  {journal} {Phys.
  Rev. B}\ }\textbf {\bibinfo {volume} {88}},\ \bibinfo {pages} {125427}
  (\bibinfo {year} {2013})}\BibitemShut {NoStop}%
\bibitem [{\citenamefont {Yang}\ and\ \citenamefont
  {Nagaosa}(2014)}]{bjYang14NC_type-TDSM}%
  \BibitemOpen
  \bibfield  {author} {\bibinfo {author} {\bibfnamefont {B.-J.}\ \bibnamefont
  {Yang}}\ and\ \bibinfo {author} {\bibfnamefont {N.}~\bibnamefont {Nagaosa}},\
  }\href {\doibase 10.1038/ncomms5898} {\bibfield  {journal} {\bibinfo
  {journal} {Nat. Commun.}\ }\textbf {\bibinfo {volume} {5}},\ \bibinfo {pages}
  {4898} (\bibinfo {year} {2014})}\BibitemShut {NoStop}%
\bibitem [{\citenamefont {Burkov}\ and\ \citenamefont
  {Kim}(2016)}]{aBurkov16PRL_TDSM-SHE}%
  \BibitemOpen
  \bibfield  {author} {\bibinfo {author} {\bibfnamefont {A.~A.}\ \bibnamefont
  {Burkov}}\ and\ \bibinfo {author} {\bibfnamefont {Y.~B.}\ \bibnamefont
  {Kim}},\ }\href {\doibase 10.1103/PhysRevLett.117.136602} {\bibfield
  {journal} {\bibinfo  {journal} {Phys. Rev. Lett.}\ }\textbf {\bibinfo
  {volume} {117}},\ \bibinfo {pages} {136602} (\bibinfo {year}
  {2016})}\BibitemShut {NoStop}%
\bibitem [{Note1()}]{Note1}%
  \BibitemOpen
  \bibinfo {note} {For the TDSM with an effective Hamiltonian~\protect \textup
  {\hbox {\mathsurround \z@ \protect \normalfont (\ignorespaces \ref
  {eqHD}\unskip \@@italiccorr )}}, the inversion and time-reversal operators
  are represented by $\protect \mathcal P = \tau _z$ and $\protect \mathcal T =
  i\sigma _y \protect \mathcal K$ (with $\protect \mathcal K$ being the complex
  conjugation operator), respectively, and $H_D(\vb k)$ satisfies $\protect
  \mathcal P^{-1} H_D(\vb k)\protect \mathcal P = H_D(-\vb k)$ and $\protect
  \mathcal T^{-1} H_D(\vb k)\protect \mathcal T = H_D(-\vb k)$.}\BibitemShut
  {Stop}%
\bibitem [{\citenamefont {Villar~Arribi}\ \emph {et~al.}(2020)\citenamefont
  {Villar~Arribi}, \citenamefont {Zhu}, \citenamefont {Schumann}, \citenamefont
  {Stemmer}, \citenamefont {Burkov},\ and\ \citenamefont
  {Heinonen}}]{Arribi20PRB_CdAs_params}%
  \BibitemOpen
  \bibfield  {author} {\bibinfo {author} {\bibfnamefont {P.}~\bibnamefont
  {Villar~Arribi}}, \bibinfo {author} {\bibfnamefont {J.-X.}\ \bibnamefont
  {Zhu}}, \bibinfo {author} {\bibfnamefont {T.}~\bibnamefont {Schumann}},
  \bibinfo {author} {\bibfnamefont {S.}~\bibnamefont {Stemmer}}, \bibinfo
  {author} {\bibfnamefont {A.~A.}\ \bibnamefont {Burkov}}, \ and\ \bibinfo
  {author} {\bibfnamefont {O.}~\bibnamefont {Heinonen}},\ }\href {\doibase
  10.1103/PhysRevB.102.155141} {\bibfield  {journal} {\bibinfo  {journal}
  {Phys. Rev. B}\ }\textbf {\bibinfo {volume} {102}},\ \bibinfo {pages}
  {155141} (\bibinfo {year} {2020})}\BibitemShut {NoStop}%
\bibitem [{\citenamefont {Villanova}\ and\ \citenamefont
  {Park}(2018)}]{Villanova18PRB_DSM_Zeeman}%
  \BibitemOpen
  \bibfield  {author} {\bibinfo {author} {\bibfnamefont {J.~W.}\ \bibnamefont
  {Villanova}}\ and\ \bibinfo {author} {\bibfnamefont {K.}~\bibnamefont
  {Park}},\ }\href {\doibase 10.1103/PhysRevB.98.075123} {\bibfield  {journal}
  {\bibinfo  {journal} {Phys. Rev. B}\ }\textbf {\bibinfo {volume} {98}},\
  \bibinfo {pages} {075123} (\bibinfo {year} {2018})}\BibitemShut {NoStop}%
\bibitem [{\citenamefont {Hashimoto}\ \emph {et~al.}(2016)\citenamefont
  {Hashimoto}, \citenamefont {Kobayashi}, \citenamefont {Tanaka},\ and\
  \citenamefont {Sato}}]{Hashimoto16PRB_sc-Zeeman-DSM}%
  \BibitemOpen
  \bibfield  {author} {\bibinfo {author} {\bibfnamefont {T.}~\bibnamefont
  {Hashimoto}}, \bibinfo {author} {\bibfnamefont {S.}~\bibnamefont
  {Kobayashi}}, \bibinfo {author} {\bibfnamefont {Y.}~\bibnamefont {Tanaka}}, \
  and\ \bibinfo {author} {\bibfnamefont {M.}~\bibnamefont {Sato}},\ }\href
  {\doibase 10.1103/PhysRevB.94.014510} {\bibfield  {journal} {\bibinfo
  {journal} {Phys. Rev. B}\ }\textbf {\bibinfo {volume} {94}},\ \bibinfo
  {pages} {014510} (\bibinfo {year} {2016})}\BibitemShut {NoStop}%
\bibitem [{\citenamefont {Jeon}\ \emph {et~al.}(2014)\citenamefont {Jeon},
  \citenamefont {Zhou}, \citenamefont {Gyenis}, \citenamefont {Feldman},
  \citenamefont {Kimchi}, \citenamefont {Potter}, \citenamefont {Gibson},
  \citenamefont {Cava}, \citenamefont {Vishwanath},\ and\ \citenamefont
  {Yazdani}}]{Jeon14_CdAs_g-factor}%
  \BibitemOpen
  \bibfield  {author} {\bibinfo {author} {\bibfnamefont {S.}~\bibnamefont
  {Jeon}}, \bibinfo {author} {\bibfnamefont {B.~B.}\ \bibnamefont {Zhou}},
  \bibinfo {author} {\bibfnamefont {A.}~\bibnamefont {Gyenis}}, \bibinfo
  {author} {\bibfnamefont {B.~E.}\ \bibnamefont {Feldman}}, \bibinfo {author}
  {\bibfnamefont {I.}~\bibnamefont {Kimchi}}, \bibinfo {author} {\bibfnamefont
  {A.~C.}\ \bibnamefont {Potter}}, \bibinfo {author} {\bibfnamefont {Q.~D.}\
  \bibnamefont {Gibson}}, \bibinfo {author} {\bibfnamefont {R.~J.}\
  \bibnamefont {Cava}}, \bibinfo {author} {\bibfnamefont {A.}~\bibnamefont
  {Vishwanath}}, \ and\ \bibinfo {author} {\bibfnamefont {A.}~\bibnamefont
  {Yazdani}},\ }\href {\doibase 10.1038/nmat4023} {\bibfield  {journal}
  {\bibinfo  {journal} {Nat. Mater.}\ }\textbf {\bibinfo {volume} {13}},\
  \bibinfo {pages} {851} (\bibinfo {year} {2014})}\BibitemShut {NoStop}%
\bibitem [{\citenamefont {Xiong}\ \emph {et~al.}(2015)\citenamefont {Xiong},
  \citenamefont {Kushwaha}, \citenamefont {Liang}, \citenamefont {Krizan},
  \citenamefont {Hirschberger}, \citenamefont {Wang}, \citenamefont {Cava},\
  and\ \citenamefont {Ong}}]{Xiong15_NaBi_g-factor}%
  \BibitemOpen
  \bibfield  {author} {\bibinfo {author} {\bibfnamefont {J.}~\bibnamefont
  {Xiong}}, \bibinfo {author} {\bibfnamefont {S.~K.}\ \bibnamefont {Kushwaha}},
  \bibinfo {author} {\bibfnamefont {T.}~\bibnamefont {Liang}}, \bibinfo
  {author} {\bibfnamefont {J.~W.}\ \bibnamefont {Krizan}}, \bibinfo {author}
  {\bibfnamefont {M.}~\bibnamefont {Hirschberger}}, \bibinfo {author}
  {\bibfnamefont {W.}~\bibnamefont {Wang}}, \bibinfo {author} {\bibfnamefont
  {R.~J.}\ \bibnamefont {Cava}}, \ and\ \bibinfo {author} {\bibfnamefont
  {N.~P.}\ \bibnamefont {Ong}},\ }\href
  {https://science.sciencemag.org/content/350/6259/413} {\bibfield  {journal}
  {\bibinfo  {journal} {Science}\ }\textbf {\bibinfo {volume} {350}},\ \bibinfo
  {pages} {413} (\bibinfo {year} {2015})}\BibitemShut {NoStop}%
\bibitem [{\citenamefont {Baidya}\ and\ \citenamefont
  {Vanderbilt}(2020)}]{Baidya20PRB_Zeeman_CdAs}%
  \BibitemOpen
  \bibfield  {author} {\bibinfo {author} {\bibfnamefont {S.}~\bibnamefont
  {Baidya}}\ and\ \bibinfo {author} {\bibfnamefont {D.}~\bibnamefont
  {Vanderbilt}},\ }\href {\doibase 10.1103/PhysRevB.102.165115} {\bibfield
  {journal} {\bibinfo  {journal} {Phys. Rev. B}\ }\textbf {\bibinfo {volume}
  {102}},\ \bibinfo {pages} {165115} (\bibinfo {year} {2020})}\BibitemShut
  {NoStop}%
\bibitem [{\citenamefont {Karplus}\ and\ \citenamefont
  {Luttinger}(1954)}]{Karplus54}%
  \BibitemOpen
  \bibfield  {author} {\bibinfo {author} {\bibfnamefont {R.}~\bibnamefont
  {Karplus}}\ and\ \bibinfo {author} {\bibfnamefont {J.~M.}\ \bibnamefont
  {Luttinger}},\ }\href {\doibase 10.1103/PhysRev.95.1154} {\bibfield
  {journal} {\bibinfo  {journal} {Phys. Rev.}\ }\textbf {\bibinfo {volume}
  {95}},\ \bibinfo {pages} {1154} (\bibinfo {year} {1954})}\BibitemShut
  {NoStop}%
\bibitem [{\citenamefont {Xiao}\ \emph {et~al.}(2010)\citenamefont {Xiao},
  \citenamefont {Chang},\ and\ \citenamefont {Niu}}]{Xiao2010}%
  \BibitemOpen
  \bibfield  {author} {\bibinfo {author} {\bibfnamefont {D.}~\bibnamefont
  {Xiao}}, \bibinfo {author} {\bibfnamefont {M.-C.}\ \bibnamefont {Chang}}, \
  and\ \bibinfo {author} {\bibfnamefont {Q.}~\bibnamefont {Niu}},\ }\href
  {\doibase 10.1103/RevModPhys.82.1959} {\bibfield  {journal} {\bibinfo
  {journal} {Rev. Mod. Phys.}\ }\textbf {\bibinfo {volume} {82}},\ \bibinfo
  {pages} {1959} (\bibinfo {year} {2010})}\BibitemShut {NoStop}%
\bibitem [{Note2()}]{Note2}%
  \BibitemOpen
  \bibinfo {note} {In general topologically trivial bands near the Fermi energy
  may also have sizable contributions to the spin Berry curvature and modify
  the spin Hall responses if band degeneracy occurs. However, in the two
  prototypical class-I TDSM materials, Cd$_3$As$_2$ and Na$_3$Bi, the pair of
  Dirac points are close to the Fermi energy with very few trivial bands
  nearby. Therefore, the low-energy electronic properties as well as the
  intrinsic SHC in these two materials are well captured by the effective
  four-band Hamiltonian \protect \textup {\hbox {\mathsurround \z@ \protect
  \normalfont (\ignorespaces \ref {eq1}\unskip \@@italiccorr )}}.}\BibitemShut
  {Stop}%
\bibitem [{\citenamefont {Burkov}(2014)}]{Burkov14PRL_AHE-WSM}%
  \BibitemOpen
  \bibfield  {author} {\bibinfo {author} {\bibfnamefont {A.~A.}\ \bibnamefont
  {Burkov}},\ }\href {\doibase 10.1103/PhysRevLett.113.187202} {\bibfield
  {journal} {\bibinfo  {journal} {Phys. Rev. Lett.}\ }\textbf {\bibinfo
  {volume} {113}},\ \bibinfo {pages} {187202} (\bibinfo {year}
  {2014})}\BibitemShut {NoStop}%
\bibitem [{\citenamefont {Hasan}\ and\ \citenamefont
  {Kane}(2010)}]{Hasan_Kane_10RMP_TI}%
  \BibitemOpen
  \bibfield  {author} {\bibinfo {author} {\bibfnamefont {M.~Z.}\ \bibnamefont
  {Hasan}}\ and\ \bibinfo {author} {\bibfnamefont {C.~L.}\ \bibnamefont
  {Kane}},\ }\href {\doibase 10.1103/RevModPhys.82.3045} {\bibfield  {journal}
  {\bibinfo  {journal} {Rev. Mod. Phys.}\ }\textbf {\bibinfo {volume} {82}},\
  \bibinfo {pages} {3045} (\bibinfo {year} {2010})}\BibitemShut {NoStop}%
\bibitem [{\citenamefont {Yao}\ \emph {et~al.}(2004)\citenamefont {Yao},
  \citenamefont {Kleinman}, \citenamefont {MacDonald}, \citenamefont {Sinova},
  \citenamefont {Jungwirth}, \citenamefont {Wang}, \citenamefont {Wang},\ and\
  \citenamefont {Niu}}]{Yao04_Kubo_AHC}%
  \BibitemOpen
  \bibfield  {author} {\bibinfo {author} {\bibfnamefont {Y.}~\bibnamefont
  {Yao}}, \bibinfo {author} {\bibfnamefont {L.}~\bibnamefont {Kleinman}},
  \bibinfo {author} {\bibfnamefont {A.~H.}\ \bibnamefont {MacDonald}}, \bibinfo
  {author} {\bibfnamefont {J.}~\bibnamefont {Sinova}}, \bibinfo {author}
  {\bibfnamefont {T.}~\bibnamefont {Jungwirth}}, \bibinfo {author}
  {\bibfnamefont {D.-s.}\ \bibnamefont {Wang}}, \bibinfo {author}
  {\bibfnamefont {E.}~\bibnamefont {Wang}}, \ and\ \bibinfo {author}
  {\bibfnamefont {Q.}~\bibnamefont {Niu}},\ }\href {\doibase
  10.1103/PhysRevLett.92.037204} {\bibfield  {journal} {\bibinfo  {journal}
  {Phys. Rev. Lett.}\ }\textbf {\bibinfo {volume} {92}},\ \bibinfo {pages}
  {037204} (\bibinfo {year} {2004})}\BibitemShut {NoStop}%
\bibitem [{\citenamefont {Dutta}\ \emph {et~al.}(2012)\citenamefont {Dutta},
  \citenamefont {Maiti},\ and\ \citenamefont {Karmakar}}]{Dutta12_Kubo_gamma}%
  \BibitemOpen
  \bibfield  {author} {\bibinfo {author} {\bibfnamefont {P.}~\bibnamefont
  {Dutta}}, \bibinfo {author} {\bibfnamefont {S.~K.}\ \bibnamefont {Maiti}}, \
  and\ \bibinfo {author} {\bibfnamefont {S.~N.}\ \bibnamefont {Karmakar}},\
  }\href {\doibase 10.1063/1.4748312} {\bibfield  {journal} {\bibinfo
  {journal} {J. Appl. Phys.}\ }\textbf {\bibinfo {volume} {112}},\ \bibinfo
  {pages} {044306} (\bibinfo {year} {2012})}\BibitemShut {NoStop}%
\bibitem [{\citenamefont {Uchida}\ \emph {et~al.}(2017)\citenamefont {Uchida},
  \citenamefont {Nakazawa}, \citenamefont {Nishihaya}, \citenamefont {Akiba},
  \citenamefont {Kriener}, \citenamefont {Kozuka}, \citenamefont {Miyake},
  \citenamefont {Taguchi}, \citenamefont {Tokunaga}, \citenamefont {Nagaosa},
  \citenamefont {Tokura},\ and\ \citenamefont {Kawasaki}}]{Uchida17_CdAs_tauq}%
  \BibitemOpen
  \bibfield  {author} {\bibinfo {author} {\bibfnamefont {M.}~\bibnamefont
  {Uchida}}, \bibinfo {author} {\bibfnamefont {Y.}~\bibnamefont {Nakazawa}},
  \bibinfo {author} {\bibfnamefont {S.}~\bibnamefont {Nishihaya}}, \bibinfo
  {author} {\bibfnamefont {K.}~\bibnamefont {Akiba}}, \bibinfo {author}
  {\bibfnamefont {M.}~\bibnamefont {Kriener}}, \bibinfo {author} {\bibfnamefont
  {Y.}~\bibnamefont {Kozuka}}, \bibinfo {author} {\bibfnamefont
  {A.}~\bibnamefont {Miyake}}, \bibinfo {author} {\bibfnamefont
  {Y.}~\bibnamefont {Taguchi}}, \bibinfo {author} {\bibfnamefont
  {M.}~\bibnamefont {Tokunaga}}, \bibinfo {author} {\bibfnamefont
  {N.}~\bibnamefont {Nagaosa}}, \bibinfo {author} {\bibfnamefont
  {Y.}~\bibnamefont {Tokura}}, \ and\ \bibinfo {author} {\bibfnamefont
  {M.}~\bibnamefont {Kawasaki}},\ }\href {\doibase 10.1038/s41467-017-02423-1}
  {\bibfield  {journal} {\bibinfo  {journal} {Nat. Commun.}\ }\textbf {\bibinfo
  {volume} {8}},\ \bibinfo {pages} {2274} (\bibinfo {year} {2017})}\BibitemShut
  {NoStop}%
\bibitem [{\citenamefont {Yanez}\ \emph {et~al.}(2021)\citenamefont {Yanez},
  \citenamefont {Ou}, \citenamefont {Xiao}, \citenamefont {Koo}, \citenamefont
  {Held}, \citenamefont {Ghosh}, \citenamefont {Rable}, \citenamefont
  {Pillsbury}, \citenamefont {Delgado}, \citenamefont {Yang}, \citenamefont
  {Chamorro}, \citenamefont {Grutter}, \citenamefont {Quarterman},
  \citenamefont {Richardella}, \citenamefont {Sengupta}, \citenamefont
  {McQueen}, \citenamefont {Borchers}, \citenamefont {Mkhoyan}, \citenamefont
  {Yan},\ and\ \citenamefont {Samarth}}]{Yanez_21_SCC_Cd3As2}%
  \BibitemOpen
  \bibfield  {author} {\bibinfo {author} {\bibfnamefont {W.}~\bibnamefont
  {Yanez}}, \bibinfo {author} {\bibfnamefont {Y.}~\bibnamefont {Ou}}, \bibinfo
  {author} {\bibfnamefont {R.}~\bibnamefont {Xiao}}, \bibinfo {author}
  {\bibfnamefont {J.}~\bibnamefont {Koo}}, \bibinfo {author} {\bibfnamefont
  {J.~T.}\ \bibnamefont {Held}}, \bibinfo {author} {\bibfnamefont
  {S.}~\bibnamefont {Ghosh}}, \bibinfo {author} {\bibfnamefont
  {J.}~\bibnamefont {Rable}}, \bibinfo {author} {\bibfnamefont
  {T.}~\bibnamefont {Pillsbury}}, \bibinfo {author} {\bibfnamefont {E.~G.}\
  \bibnamefont {Delgado}}, \bibinfo {author} {\bibfnamefont {K.}~\bibnamefont
  {Yang}}, \bibinfo {author} {\bibfnamefont {J.}~\bibnamefont {Chamorro}},
  \bibinfo {author} {\bibfnamefont {A.~J.}\ \bibnamefont {Grutter}}, \bibinfo
  {author} {\bibfnamefont {P.}~\bibnamefont {Quarterman}}, \bibinfo {author}
  {\bibfnamefont {A.}~\bibnamefont {Richardella}}, \bibinfo {author}
  {\bibfnamefont {A.}~\bibnamefont {Sengupta}}, \bibinfo {author}
  {\bibfnamefont {T.}~\bibnamefont {McQueen}}, \bibinfo {author} {\bibfnamefont
  {J.~A.}\ \bibnamefont {Borchers}}, \bibinfo {author} {\bibfnamefont {K.~A.}\
  \bibnamefont {Mkhoyan}}, \bibinfo {author} {\bibfnamefont {B.}~\bibnamefont
  {Yan}}, \ and\ \bibinfo {author} {\bibfnamefont {N.}~\bibnamefont
  {Samarth}},\ }\href {\doibase 10.1103/PhysRevApplied.16.054031} {\bibfield
  {journal} {\bibinfo  {journal} {Phys. Rev. Applied}\ }\textbf {\bibinfo
  {volume} {16}},\ \bibinfo {pages} {054031} (\bibinfo {year}
  {2021})}\BibitemShut {NoStop}%
\bibitem [{\citenamefont {Liang}\ \emph {et~al.}(2015)\citenamefont {Liang},
  \citenamefont {Gibson}, \citenamefont {Ali}, \citenamefont {Liu},
  \citenamefont {Cava},\ and\ \citenamefont {Ong}}]{liang2015}%
  \BibitemOpen
  \bibfield  {author} {\bibinfo {author} {\bibfnamefont {T.}~\bibnamefont
  {Liang}}, \bibinfo {author} {\bibfnamefont {Q.}~\bibnamefont {Gibson}},
  \bibinfo {author} {\bibfnamefont {M.~N.}\ \bibnamefont {Ali}}, \bibinfo
  {author} {\bibfnamefont {M.}~\bibnamefont {Liu}}, \bibinfo {author}
  {\bibfnamefont {R.~J.}\ \bibnamefont {Cava}}, \ and\ \bibinfo {author}
  {\bibfnamefont {N.~P.}\ \bibnamefont {Ong}},\ }\href {\doibase
  10.1038/nmat4143} {\bibfield  {journal} {\bibinfo  {journal} {Nat. Mater.}\
  }\textbf {\bibinfo {volume} {14}},\ \bibinfo {pages} {280} (\bibinfo {year}
  {2015})}\BibitemShut {NoStop}%
\bibitem [{\citenamefont {Li}\ \emph {et~al.}(2016)\citenamefont {Li},
  \citenamefont {He}, \citenamefont {Lu}, \citenamefont {Zhang}, \citenamefont
  {Liu}, \citenamefont {Ma}, \citenamefont {Fan}, \citenamefont {Shen},\ and\
  \citenamefont {Wang}}]{li2016}%
  \BibitemOpen
  \bibfield  {author} {\bibinfo {author} {\bibfnamefont {H.}~\bibnamefont
  {Li}}, \bibinfo {author} {\bibfnamefont {H.}~\bibnamefont {He}}, \bibinfo
  {author} {\bibfnamefont {H.-Z.}\ \bibnamefont {Lu}}, \bibinfo {author}
  {\bibfnamefont {H.}~\bibnamefont {Zhang}}, \bibinfo {author} {\bibfnamefont
  {H.}~\bibnamefont {Liu}}, \bibinfo {author} {\bibfnamefont {R.}~\bibnamefont
  {Ma}}, \bibinfo {author} {\bibfnamefont {Z.}~\bibnamefont {Fan}}, \bibinfo
  {author} {\bibfnamefont {S.-Q.}\ \bibnamefont {Shen}}, \ and\ \bibinfo
  {author} {\bibfnamefont {J.}~\bibnamefont {Wang}},\ }\href {\doibase
  10.1038/ncomms10301} {\bibfield  {journal} {\bibinfo  {journal} {Nat.
  Commun.}\ }\textbf {\bibinfo {volume} {7}},\ \bibinfo {pages} {10301}
  (\bibinfo {year} {2016})}\BibitemShut {NoStop}%
\bibitem [{\citenamefont {Li}\ \emph {et~al.}(2018)\citenamefont {Li},
  \citenamefont {Wang}, \citenamefont {He}, \citenamefont {Wang},\ and\
  \citenamefont {Shen}}]{li2018}%
  \BibitemOpen
  \bibfield  {author} {\bibinfo {author} {\bibfnamefont {H.}~\bibnamefont
  {Li}}, \bibinfo {author} {\bibfnamefont {H.-W.}\ \bibnamefont {Wang}},
  \bibinfo {author} {\bibfnamefont {H.}~\bibnamefont {He}}, \bibinfo {author}
  {\bibfnamefont {J.}~\bibnamefont {Wang}}, \ and\ \bibinfo {author}
  {\bibfnamefont {S.-Q.}\ \bibnamefont {Shen}},\ }\href {\doibase
  10.1103/PhysRevB.97.201110} {\bibfield  {journal} {\bibinfo  {journal} {Phys.
  Rev. B}\ }\textbf {\bibinfo {volume} {97}},\ \bibinfo {pages} {201110}
  (\bibinfo {year} {2018})}\BibitemShut {NoStop}%
\bibitem [{\citenamefont {Nishihaya}\ \emph {et~al.}(2018)\citenamefont
  {Nishihaya}, \citenamefont {Uchida}, \citenamefont {Nakazawa}, \citenamefont
  {Kriener}, \citenamefont {Kozuka}, \citenamefont {Taguchi},\ and\
  \citenamefont {Kawasaki}}]{Nishihaya2018Sci}%
  \BibitemOpen
  \bibfield  {author} {\bibinfo {author} {\bibfnamefont {S.}~\bibnamefont
  {Nishihaya}}, \bibinfo {author} {\bibfnamefont {M.}~\bibnamefont {Uchida}},
  \bibinfo {author} {\bibfnamefont {Y.}~\bibnamefont {Nakazawa}}, \bibinfo
  {author} {\bibfnamefont {M.}~\bibnamefont {Kriener}}, \bibinfo {author}
  {\bibfnamefont {Y.}~\bibnamefont {Kozuka}}, \bibinfo {author} {\bibfnamefont
  {Y.}~\bibnamefont {Taguchi}}, \ and\ \bibinfo {author} {\bibfnamefont
  {M.}~\bibnamefont {Kawasaki}},\ }\href {\doibase 10.1126/sciadv.aar5668}
  {\bibfield  {journal} {\bibinfo  {journal} {Sci. Adv.}\ }\textbf {\bibinfo
  {volume} {4}},\ \bibinfo {pages} {eaar5668} (\bibinfo {year}
  {2018})}\BibitemShut {NoStop}%
\bibitem [{\citenamefont {Zhang}\ \emph
  {et~al.}(2017{\natexlab{a}})\citenamefont {Zhang}, \citenamefont {Sun},
  \citenamefont {Liu}, \citenamefont {Narayan}, \citenamefont {Li},
  \citenamefont {Yuan}, \citenamefont {Liu}, \citenamefont {Dai}, \citenamefont
  {Long}, \citenamefont {Uwatoko}, \citenamefont {Shen}, \citenamefont
  {Sanvito}, \citenamefont {Yang}, \citenamefont {Cheng},\ and\ \citenamefont
  {Xiu}}]{zhang2017pressure}%
  \BibitemOpen
  \bibfield  {author} {\bibinfo {author} {\bibfnamefont {C.}~\bibnamefont
  {Zhang}}, \bibinfo {author} {\bibfnamefont {J.}~\bibnamefont {Sun}}, \bibinfo
  {author} {\bibfnamefont {F.}~\bibnamefont {Liu}}, \bibinfo {author}
  {\bibfnamefont {A.}~\bibnamefont {Narayan}}, \bibinfo {author} {\bibfnamefont
  {N.}~\bibnamefont {Li}}, \bibinfo {author} {\bibfnamefont {X.}~\bibnamefont
  {Yuan}}, \bibinfo {author} {\bibfnamefont {Y.}~\bibnamefont {Liu}}, \bibinfo
  {author} {\bibfnamefont {J.}~\bibnamefont {Dai}}, \bibinfo {author}
  {\bibfnamefont {Y.}~\bibnamefont {Long}}, \bibinfo {author} {\bibfnamefont
  {Y.}~\bibnamefont {Uwatoko}}, \bibinfo {author} {\bibfnamefont
  {J.}~\bibnamefont {Shen}}, \bibinfo {author} {\bibfnamefont {S.}~\bibnamefont
  {Sanvito}}, \bibinfo {author} {\bibfnamefont {W.}~\bibnamefont {Yang}},
  \bibinfo {author} {\bibfnamefont {J.}~\bibnamefont {Cheng}}, \ and\ \bibinfo
  {author} {\bibfnamefont {F.}~\bibnamefont {Xiu}},\ }\href {\doibase
  10.1103/PhysRevB.96.155205} {\bibfield  {journal} {\bibinfo  {journal} {Phys.
  Rev. B}\ }\textbf {\bibinfo {volume} {96}},\ \bibinfo {pages} {155205}
  (\bibinfo {year} {2017}{\natexlab{a}})}\BibitemShut {NoStop}%
\bibitem [{Note3()}]{Note3}%
  \BibitemOpen
  \bibinfo {note} {Some literature studies have called them type-II Dirac
  semimetals. However, the name ``type-II'' in the context of Weyl semimetals
  usually refers to Weyl semimetals with over-tilted Weyl cones. Therefore, to
  avoid unnecessary confusion, we refrain from such nomenclature and refer to
  the Dirac semimetals with single Dirac points located at the TRIM as class-II
  Dirac semimetals, in contrast to class-I considered in our study where the
  Dirac points are located on the rotation axis but not pinned to the
  TRIM.}\BibitemShut {Stop}%
\bibitem [{\citenamefont {Burkov}(2018)}]{Burkov18_mirror_anomaly_DSM}%
  \BibitemOpen
  \bibfield  {author} {\bibinfo {author} {\bibfnamefont {A.~A.}\ \bibnamefont
  {Burkov}},\ }\href {\doibase 10.1103/PhysRevLett.120.016603} {\bibfield
  {journal} {\bibinfo  {journal} {Phys. Rev. Lett.}\ }\textbf {\bibinfo
  {volume} {120}},\ \bibinfo {pages} {016603} (\bibinfo {year}
  {2018})}\BibitemShut {NoStop}%
\bibitem [{\citenamefont {Nandy}\ \emph {et~al.}(2019)\citenamefont {Nandy},
  \citenamefont {Saha}, \citenamefont {Taraphder},\ and\ \citenamefont
  {Tewari}}]{Nandy19PRB_MA_typeI_DSM}%
  \BibitemOpen
  \bibfield  {author} {\bibinfo {author} {\bibfnamefont {S.}~\bibnamefont
  {Nandy}}, \bibinfo {author} {\bibfnamefont {K.}~\bibnamefont {Saha}},
  \bibinfo {author} {\bibfnamefont {A.}~\bibnamefont {Taraphder}}, \ and\
  \bibinfo {author} {\bibfnamefont {S.}~\bibnamefont {Tewari}},\ }\href
  {\doibase 10.1103/PhysRevB.99.075116} {\bibfield  {journal} {\bibinfo
  {journal} {Phys. Rev. B}\ }\textbf {\bibinfo {volume} {99}},\ \bibinfo
  {pages} {075116} (\bibinfo {year} {2019})}\BibitemShut {NoStop}%
\bibitem [{\citenamefont {Zyuzin}\ and\ \citenamefont
  {Tiwari}(2016)}]{Zyuzin2016}%
  \BibitemOpen
  \bibfield  {author} {\bibinfo {author} {\bibfnamefont {A.~A.}\ \bibnamefont
  {Zyuzin}}\ and\ \bibinfo {author} {\bibfnamefont {R.~P.}\ \bibnamefont
  {Tiwari}},\ }\href {\doibase 10.1134/S002136401611014X} {\bibfield  {journal}
  {\bibinfo  {journal} {JETP Lett.}\ }\textbf {\bibinfo {volume} {103}},\
  \bibinfo {pages} {717} (\bibinfo {year} {2016})}\BibitemShut {NoStop}%
\bibitem [{\citenamefont {Liu}\ \emph {et~al.}(2010)\citenamefont {Liu},
  \citenamefont {Qi}, \citenamefont {Zhang}, \citenamefont {Dai}, \citenamefont
  {Fang},\ and\ \citenamefont {Zhang}}]{Liu10PRB_model_H_TI}%
  \BibitemOpen
  \bibfield  {author} {\bibinfo {author} {\bibfnamefont {C.-X.}\ \bibnamefont
  {Liu}}, \bibinfo {author} {\bibfnamefont {X.-L.}\ \bibnamefont {Qi}},
  \bibinfo {author} {\bibfnamefont {H.}~\bibnamefont {Zhang}}, \bibinfo
  {author} {\bibfnamefont {X.}~\bibnamefont {Dai}}, \bibinfo {author}
  {\bibfnamefont {Z.}~\bibnamefont {Fang}}, \ and\ \bibinfo {author}
  {\bibfnamefont {S.-C.}\ \bibnamefont {Zhang}},\ }\href {\doibase
  10.1103/PhysRevB.82.045122} {\bibfield  {journal} {\bibinfo  {journal} {Phys.
  Rev. B}\ }\textbf {\bibinfo {volume} {82}},\ \bibinfo {pages} {045122}
  (\bibinfo {year} {2010})}\BibitemShut {NoStop}%
\bibitem [{\citenamefont {Potter}\ \emph {et~al.}(2014)\citenamefont {Potter},
  \citenamefont {Kimchi},\ and\ \citenamefont
  {Vishwanath}}]{Potter14ncomm_QO_TSM}%
  \BibitemOpen
  \bibfield  {author} {\bibinfo {author} {\bibfnamefont {A.~C.}\ \bibnamefont
  {Potter}}, \bibinfo {author} {\bibfnamefont {I.}~\bibnamefont {Kimchi}}, \
  and\ \bibinfo {author} {\bibfnamefont {A.}~\bibnamefont {Vishwanath}},\
  }\href {\doibase 10.1038/ncomms6161} {\bibfield  {journal} {\bibinfo
  {journal} {Nat. Commun.}\ }\textbf {\bibinfo {volume} {5}},\ \bibinfo {pages}
  {5161} (\bibinfo {year} {2014})}\BibitemShut {NoStop}%
\bibitem [{\citenamefont {Zhang}\ \emph
  {et~al.}(2017{\natexlab{b}})\citenamefont {Zhang}, \citenamefont {Narayan},
  \citenamefont {Lu}, \citenamefont {Zhang}, \citenamefont {Zhang},
  \citenamefont {Ni}, \citenamefont {Yuan}, \citenamefont {Liu}, \citenamefont
  {Park}, \citenamefont {Zhang}, \citenamefont {Wang}, \citenamefont {Liu},
  \citenamefont {Cheng}, \citenamefont {Pi}, \citenamefont {Sheng},
  \citenamefont {Sanvito},\ and\ \citenamefont
  {Xiu}}]{Zhang17ncomm_Weyl_orbit_DSM}%
  \BibitemOpen
  \bibfield  {author} {\bibinfo {author} {\bibfnamefont {C.}~\bibnamefont
  {Zhang}}, \bibinfo {author} {\bibfnamefont {A.}~\bibnamefont {Narayan}},
  \bibinfo {author} {\bibfnamefont {S.}~\bibnamefont {Lu}}, \bibinfo {author}
  {\bibfnamefont {J.}~\bibnamefont {Zhang}}, \bibinfo {author} {\bibfnamefont
  {H.}~\bibnamefont {Zhang}}, \bibinfo {author} {\bibfnamefont
  {Z.}~\bibnamefont {Ni}}, \bibinfo {author} {\bibfnamefont {X.}~\bibnamefont
  {Yuan}}, \bibinfo {author} {\bibfnamefont {Y.}~\bibnamefont {Liu}}, \bibinfo
  {author} {\bibfnamefont {J.-H.}\ \bibnamefont {Park}}, \bibinfo {author}
  {\bibfnamefont {E.}~\bibnamefont {Zhang}}, \bibinfo {author} {\bibfnamefont
  {W.}~\bibnamefont {Wang}}, \bibinfo {author} {\bibfnamefont {S.}~\bibnamefont
  {Liu}}, \bibinfo {author} {\bibfnamefont {L.}~\bibnamefont {Cheng}}, \bibinfo
  {author} {\bibfnamefont {L.}~\bibnamefont {Pi}}, \bibinfo {author}
  {\bibfnamefont {Z.}~\bibnamefont {Sheng}}, \bibinfo {author} {\bibfnamefont
  {S.}~\bibnamefont {Sanvito}}, \ and\ \bibinfo {author} {\bibfnamefont
  {F.}~\bibnamefont {Xiu}},\ }\href {\doibase 10.1038/s41467-017-01438-y}
  {\bibfield  {journal} {\bibinfo  {journal} {Nat. Commun.}\ }\textbf {\bibinfo
  {volume} {8}},\ \bibinfo {pages} {1272} (\bibinfo {year}
  {2017}{\natexlab{b}})}\BibitemShut {NoStop}%
\bibitem [{\citenamefont {Zhang}\ \emph
  {et~al.}(2019{\natexlab{b}})\citenamefont {Zhang}, \citenamefont {Zhang},
  \citenamefont {Yuan}, \citenamefont {Lu}, \citenamefont {Zhang},
  \citenamefont {Narayan}, \citenamefont {Liu}, \citenamefont {Zhang},
  \citenamefont {Ni}, \citenamefont {Liu}, \citenamefont {Choi}, \citenamefont
  {Suslov}, \citenamefont {Sanvito}, \citenamefont {Pi}, \citenamefont {Lu},
  \citenamefont {Potter},\ and\ \citenamefont
  {Xiu}}]{Zhang19nature_QHE_Cd3As2}%
  \BibitemOpen
  \bibfield  {author} {\bibinfo {author} {\bibfnamefont {C.}~\bibnamefont
  {Zhang}}, \bibinfo {author} {\bibfnamefont {Y.}~\bibnamefont {Zhang}},
  \bibinfo {author} {\bibfnamefont {X.}~\bibnamefont {Yuan}}, \bibinfo {author}
  {\bibfnamefont {S.}~\bibnamefont {Lu}}, \bibinfo {author} {\bibfnamefont
  {J.}~\bibnamefont {Zhang}}, \bibinfo {author} {\bibfnamefont
  {A.}~\bibnamefont {Narayan}}, \bibinfo {author} {\bibfnamefont
  {Y.}~\bibnamefont {Liu}}, \bibinfo {author} {\bibfnamefont {H.}~\bibnamefont
  {Zhang}}, \bibinfo {author} {\bibfnamefont {Z.}~\bibnamefont {Ni}}, \bibinfo
  {author} {\bibfnamefont {R.}~\bibnamefont {Liu}}, \bibinfo {author}
  {\bibfnamefont {E.~S.}\ \bibnamefont {Choi}}, \bibinfo {author}
  {\bibfnamefont {A.}~\bibnamefont {Suslov}}, \bibinfo {author} {\bibfnamefont
  {S.}~\bibnamefont {Sanvito}}, \bibinfo {author} {\bibfnamefont
  {L.}~\bibnamefont {Pi}}, \bibinfo {author} {\bibfnamefont {H.-Z.}\
  \bibnamefont {Lu}}, \bibinfo {author} {\bibfnamefont {A.~C.}\ \bibnamefont
  {Potter}}, \ and\ \bibinfo {author} {\bibfnamefont {F.}~\bibnamefont {Xiu}},\
  }\href {\doibase 10.1038/s41586-018-0798-3} {\bibfield  {journal} {\bibinfo
  {journal} {Nature}\ }\textbf {\bibinfo {volume} {565}},\ \bibinfo {pages}
  {331} (\bibinfo {year} {2019}{\natexlab{b}})}\BibitemShut {NoStop}%
\bibitem [{\citenamefont {Wawrzik}\ \emph {et~al.}(2021)\citenamefont
  {Wawrzik}, \citenamefont {You}, \citenamefont {Facio}, \citenamefont {van~den
  Brink},\ and\ \citenamefont {Sodemann}}]{Wawrzik21PRL_hot_line_WSM}%
  \BibitemOpen
  \bibfield  {author} {\bibinfo {author} {\bibfnamefont {D.}~\bibnamefont
  {Wawrzik}}, \bibinfo {author} {\bibfnamefont {J.-S.}\ \bibnamefont {You}},
  \bibinfo {author} {\bibfnamefont {J.~I.}\ \bibnamefont {Facio}}, \bibinfo
  {author} {\bibfnamefont {J.}~\bibnamefont {van~den Brink}}, \ and\ \bibinfo
  {author} {\bibfnamefont {I.}~\bibnamefont {Sodemann}},\ }\href {\doibase
  10.1103/PhysRevLett.127.056601} {\bibfield  {journal} {\bibinfo  {journal}
  {Phys. Rev. Lett.}\ }\textbf {\bibinfo {volume} {127}},\ \bibinfo {pages}
  {056601} (\bibinfo {year} {2021})}\BibitemShut {NoStop}%
\bibitem [{\citenamefont {Utermohlen}\ and\ \citenamefont
  {Trivedi}(2021)}]{Utermohlen21PRB_tensor}%
  \BibitemOpen
  \bibfield  {author} {\bibinfo {author} {\bibfnamefont {F.~G.}\ \bibnamefont
  {Utermohlen}}\ and\ \bibinfo {author} {\bibfnamefont {N.}~\bibnamefont
  {Trivedi}},\ }\href {\doibase 10.1103/PhysRevB.103.155124} {\bibfield
  {journal} {\bibinfo  {journal} {Phys. Rev. B}\ }\textbf {\bibinfo {volume}
  {103}},\ \bibinfo {pages} {155124} (\bibinfo {year} {2021})}\BibitemShut
  {NoStop}%
\end{thebibliography}%

\end{document}